# Dwarf Elliptical Galaxies

*Henry C. Ferguson*[1]
Space Telescope Science Institute [2]
3700 San Martin Drive, Baltimore, MD 21218
ferguson@stsci.edu

*Bruno Binggeli*
Astronomisches Institut der Universität Basel
Venusstrasse 7, CH-4102 Binningen, Switzerland
binggeli2@urz.unibas.ch

astro-ph/9409079  28 Sep 94

# ABSTRACT

Dwarf elliptical (dE) galaxies, with blue absolute magnitudes typically fainter than $M_B = -16$, are the most numerous type of galaxy in the nearby universe. Tremendous advances have been made over the past several years in delineating the properties of both Local Group satellite dE's and the large dE populations of nearby clusters. We review some of these advances, with particular attention to how well currently available data can constrain

(a) models for the formation of dE's,

(b) the physical and evolutionary connections between different types of galaxies (nucleated and nonnucleated dE's, compact E's, irregulars, and blue compact dwarfs) that overlap in the same portion of the mass-spectrum of galaxies,

(c) the contribution of dE's to the galaxy luminosity functions in clusters and the field,

(d) the star-forming histories of dE's and their possible contribution to faint galaxy counts, and

(e) the clustering properties of dE's.

In addressing these issues, we highlight the extent to which selection effects temper these constraints, and outline areas where new data would be particularly valuable.

# 1. Introduction

The properties of dwarf galaxies are central to two current issues in observational cosmology: understanding the results from deep redshift surveys, and interpreting the constraints of the galaxy luminosity function on the primordial fluctuation spectrum. Deep redshift surveys have revealed a dominant population of star-forming dwarf galaxies at redshifts ($0.2 < z < 0.5$) (Broadhurst et al. 1988; Colless et al. 1990; Cowie et al. 1991). A large dwarf population is *expected* from simulations of galaxy evolution (White and Frenk 1991) based on the Cold-Dark-Matter (CDM) model. However, both the redshift surveys and the theory predict far more dwarfs than are observed locally. This puzzle points to a need to understand better the physics of dwarf galaxies.

The last decade has seen a surge of activity in the study of dwarf galaxies, inspired in part by such cosmological issues, but more directly by the availability of new technology: wide-field imaging telescopes and good photographic emulsions, CCD's, and sensitive spectrographs. This activity has resulted in major advances in our knowledge of the internal structure and kinematics, stellar populations, clustering properties, luminosity function, and evolution of dwarf galaxies. Our aim in this review is to examine some of the recent data with the cosmological issues in mind: do current observations of dwarf galaxies really constrain the theories? What future observations are most relevant to testing the models of galaxy evolution? To make the subject tractable, we have chosen to concentrate on dwarf elliptical (dE) galaxies, linking together both Local Group satellite dE's and the large dE populations of nearby clusters, and examining the data critically to try to decide what we know and what we only *think* we know about these galaxies. We will mention only in passing star-forming dwarf galaxies such as Magellanic irregulars or blue compact dwarf galaxies. Excellent reviews of the properties of these galaxies can be found in Kunth et al. (1986), Hunter & Gallagher (1989), and Meylan & Prugniel (1994).

## 1.1. What is a dE Galaxy?

Low-luminosity elliptical galaxies are distinguished from late-type galaxies (spirals and irregulars) by their smooth surface-brightness profiles. Below luminosities of $M_B \approx -18$ the smooth-profile galaxies divide into two classes: compact galaxies with high central surface brightnesses (exemplified by M32), and diffuse galaxies with low central surface brightnesses (exemplified by the Local Group dwarf spheroidals). The terms "dwarf elliptical" (dE) and "dwarf spheroidal" (dSph) have been used most often to describe smooth, low surface brightness (LSB) galaxies. However, the lack of a universally accepted definition has led to some confusion over whether these terms refer to the same thing, and in particular whether the dE class includes galaxies like M32. In the discussions that follow, we adopt the classification scheme set out in the extensive Virgo cluster dwarf atlas of Sandage & Binggeli (1984). In this scheme the term dE encompasses both local dSph galaxies and similar-looking galaxies beyond the Local Group. Faint ellipticals with profiles that are more nearly $r^{1/4}$-law are referred to simply as "ellipticals" (E), or sometimes "compact ellipticals," but never dE's. A different name for "dwarf elliptical" frequently encountered in the literature is "spheroidal". The pros and cons of either name convention are discussed in Binggeli (1994b) and Kormendy

& Bender (1994).

More detailed issues of morphology, including the quantitative distinctions between the different classes and the existence of intermediate types, are discussed in §2.1.

All distance-dependent quantities discussed in this paper are based on a long distance scale ($H_0 \approx 50 \mathrm{\,km\,s^{-1} Mpc^{-1}}$), although in most cases the distances are not estimated from radial velocities.

### 1.2. A brief history

Studies of dwarf elliptical galaxies began with Shapley's discovery of the Fornax and Sculptor dE companions to the Milky Way (Shapley 1938a). At that time it was suspected that more luminous examples of such systems might exist in the Virgo cluster (Shapley 1938b), and similar companions to M31 were soon identified. Baade (1944a; 1944b) and Shapley were able to resolve the brightest stars in Local Group dwarf spheroidals and identify these systems as "population II" objects, lacking the luminous OB stars present in the galactic disk. While these systems were similar to globular clusters and elliptical galaxies in their stellar populations and smooth structure, the low surface density of stars prompted Baade (1944b) to remark that "a marked change in the internal structure of E-type nebulae takes place as we reach the systems of the lowest luminosity. The strong concentration toward the center and the central nucleus disappear gradually until we encounter such limiting forms as the Sculptor and Fornax Systems." From the numbers of Local Group dwarfs known in 1944, Baade also noted that "there seems little doubt that the symmetrical form of the luminosity function hitherto adopted has to be replaced with a skew distribution." Baade's conclusions about the structure and luminosity function of low-luminosity elliptical galaxies are echoed in modern studies of the dE populations in clusters of galaxies.

Identification of more Local Group dwarfs continued with the advent of the Palomar Sky Survey (Harrington & Wilson 1950; Wilson 1955). Searches for low-surface-brightness objects yielded similar objects outside the Local Group (Van den Bergh 1959; 1972), often concentrated around a nearby giant galaxy (Holmberg 1950; 1969). In parallel with the early identifications of nearby dwarf spheroidals and dwarf irregulars, evidence accumulated for a significant population of low-surface-brightness dwarf galaxies in nearby clusters (Shapley 1943; Reaves 1956; Hodge 1959; Hodge 1960; Hodge et al. 1965). Their spatial distribution relative to the surrounding bright galaxies immediately implicated these faint, diffuse galaxies as cluster members. However, because of their low surface-brightnesses, the systematic properties of cluster dwarfs remained largely unexplored until large-scale photographic surveys of nearby clusters allowed a detailed examination of their spatial distribution, luminosity function, and structural scaling relations (e.g. Binggeli et al. 1985; Caldwell 1987; Ferguson & Sandage 1988), and spectroscopy and multi-color observations began to reveal the properties of their stellar populations (Bothun & Caldwell 1984; Bothun et al. 1985).

The inventory of Milky Way satellites now appears virtually complete, at least out of the plane and down to luminosities $M_B \approx -9$ (Irwin 1994). Because of the difficulty of obtaining redshifts, catalogues of dE galaxies in other environments are less complete and more prone to contamination. Companions of galaxies within $\sim 5$ Mpc have been identified in various studies (see Table 1), but the physical association is in many cases uncertain.

The problem of contamination is less severe in the richest local environments (the Virgo and Fornax clusters), as the contrast of these clusters with the surrounding background is much higher and the clusters are close enough that dE galaxies can be readily distinguished from background galaxies. The principal difficulty in these surveys is incompleteness, both at low surface brightnesses due to the sensitivity of photographic plates (Impey et al. 1988) and at the high surface brightnesses due to confusion with background objects. At velocities more than about twice that of Virgo, the task of identifying dE galaxies becomes much more difficult, and contamination by background galaxies is again a concern. By distances of $\sim 5000$ km s$^{-1}$, the separation of cluster and background is best done statistically by comparison with blank fields. With the exception of Coma (Thompson & Gregory 1993), the dE population in rich clusters is essentially unknown. Populations of dE galaxies have been identified in a few loose groups with $v < 2700$ km s$^{-1}$ (Ferguson & Sandage 1991), and as companions of bright elliptical galaxies (Vader & Sandage 1991). Only a few examples of isolated dE galaxies are known, but the selection biases against finding such galaxies are severe. The clustering properties of dE galaxies are discussed in §6.

### 1.3. Theory

Theories for the formation and evolution of dE galaxies are reviewed in §7. However, to motivate the discussion of the observations, it is useful first to describe briefly the theoretical milieu.

While they appear to be the natural low-luminosity extension of the E galaxy family, the structural differences between giant and dwarf E's suggest that different physical processes govern the evolution of each type. The standard picture is that dwarf galaxies, like giants, formed from the gravitational collapse of primordial density fluctuations. In hierarchical models, once the cosmological parameters and power spectrum are specified, the evolution of dissipationless dark-matter halos can be calculated (e.g. following Press & Schechter 1974). However, the emergence of galaxies as *separate entities* in the clustering hierarchy depends on the ability of the baryons in a given overdense region to cool and form stars. The requirement that the cooling time be less than the dynamical time sets a natural *upper* limit to the galaxy mass function (Binney 1977; Rees & Ostriker 1977; Silk 1977; White & Rees 1978). Once the baryons cool and start to form stars, feedback of energy into the interstellar medium is likely to regulate (or even halt) any subsequent star formation, and mass loss may modify the galaxies' structure. Originally proposed as a mechanism for clearing gas from giant elliptical galaxies (Burke 1968; Matthews & Baker 1971), winds from supernovae were soon recognized as potentially important in governing both the structure and stellar populations of low-luminosity ellipticals (Larson 1974). Models that invoke the cessation of star formation by supernova-driven winds provide a plausible explanation for the variation of density (surface brightness) and metallicity (color) with luminosity (Larson 1974; Saito 1979; Vader 1986; Dekel & Silk 1986; Arimoto & Yoshii 1987). The status of the observed correlations between color, metallicity, and luminosity are reviewed in §4. Such models also predict high values of $M/L$ for low-luminosity galaxies; constraints from Local Group dE's will be discussed in §3.

Hierarchical models predict a power law for the low-mass tail of galaxy halos. Simple scaling relations between mass and star-formation efficiency typically also predict a power law, with a different exponent, for the faint end of the galaxy luminosity function (White & Rees 1978; White & Frenk 1991). Because dwarf galaxies condense from smaller perturbations than giants, models based on Gaussian random-phase fluctuations also predict that dwarfs will be less clustered than giants (Dekel & Silk 1986; White et al. 1987). The luminosity function and clustering properties of dwarf galaxies are thus of great interest for testing cosmological models, and will be reviewed in §5 and §6.

Table 1: Selected Catalogs of dE Galaxies

| Environment | Number of dE's | Reference |
|---|---|---|
| Local Group | 15 | Hodge 1994 |
| M81 Group | 6 | Börngen et al. 1982 |
| | | Karachentseva *et al.* 1985 |
| Bright (RSA) Galaxy Companions | 500 | Vader & Sandage 1994 |
| Nearby field (CnV & U Ma Clouds) | 50 | Binggeli *et al.* 1990 |
| Northern LSB galaxies | 20 | Schombert & Bothun 1988 |
| | | Schombert *et al.* 1992 |
| | | [cf. also Nilson 1973] |
| Southern field dwarfs | 50 | Feitzinger & Galinsky 1985 |
| Leo, Dorado, Antlia, | 350 | Ferguson & Sandage 1990 |
| Eridanus, & NGC5044 Groups | | |
| Virgo cluster | 900 | Reaves 1983 |
| | | Binggeli *et al.* 1985 |
| | | Impey *et al.* 1988 |
| Fornax cluster | 250 | Caldwell 1987 |
| | | Ferguson 1989 |
| | | Irwin et al. 1990 |
| | | Bothun et al. 1991 |
| Centaurus cluster | 200 | Bothun *et al.* 1989 |
| | | Jerjen 1994 |
| Coma cluster | 700 | Thompson & Gregory 1993 |
| | | [catalog unpublished] |

While models involving supernova-driven winds have been the most thoroughly developed, other processes may be important — or even dominant — in governing dE evolution. Other ways to remove gas include stripping by a nearby galaxy corona (Einasto et al. 1974; Lin & Faber 1983; Kormendy 1986), or sweeping by an intracluster medium. Such mechanisms predict an environmental dependence of dwarf galaxy properties that would not be expected if internal feedback were always the dominant regulator of star formation. It is also possible that star-formation is triggered in dwarf galaxies by tidal interaction with a neighbor (Lacey & Silk 1991), shocks caused by interaction with intergalactic gas in groups or clusters (Silk et al. 1987), or variations in the ionizing UV radiation field (Babul & Rees 1992; Efstathiou 1992). Triggered star formation at late epochs ($0.3 < z < 1.0$) could help account for the apparent excess of low-luminosity galaxies in deep redshift surveys. In §4 we will consider the constraints set on such models from observations of the stellar populations in *nearby* dwarfs.

Alternatively, or perhaps in addition to these other effects, it is possible that starformation in dwarfs is a cyclical process as gas repeatedly cools to form stars, then gets reheated (but not completely ejected) by OB stars and supernovae (Gerola et al. 1980; Lin & Murray 1992). Finally, there are suggestions that some dwarfs form as debris from either the explosion (during an initial burst of star formation) or collision of galaxies (Gerola et al. 1983; Barnes & Hernquist 1992; Mirabel et al. 1992). Such models are clearly radically different from the collapse onto primordial density perturbations, and expectations for the scaling relations between color, luminosity, surface brightness, and velocity dispersion remain to be worked out. Nevertheless, formation of at least some dwarfs as debris from collisions may help explain the large number of dwarfs in clusters (§6.2) and the scatter in the color-magnitude relation (§4).

## 2. Structure

### 2.1. Morphological Distinction

Dwarf ellipticals span a range of at least $10^4$ in luminosity (from $M_B \approx -18$ to $-8$) along a sequence of increasing mean surface brightness with increasing luminosity (see §2.2.2). At the bright end of this sequence, the relatively high surface brightness of a dE can mimic a normal elliptical, and quantitative analysis of surface brightness profiles is necessary to attempt a dE versus E distinction. This has proven difficult (see §2.2.1). From a purely morphological (classificatory) point of view, there will always be a certain number of bright "intermediate" (E/dE) types, irrespective of a possible discontinuity in various measurable properties between E's and dE's (Binggeli & Cameron 1991; Prugniel 1994; Vader & Chaboyer 1994).

A similar problem exists for the dE versus Irr distinction at the faint end; in fact, for the whole range $M_B < -16$. Dwarf irregulars in this range can appear very smooth and dE-like – presumably when they happen to be "sleeping," i.e. at a low or zero star formation rate. In the Virgo cluster, there is a broad dE/Irr class of galaxies comprising roughly 10% of the whole dwarf population, with increasing percentage faintwards (Sandage & Binggeli 1984; Sandage et al. 1985b). It has become clear over the years that this is not just a problem of classification (as it might be in the case of E versus dE): there appears to be a continuum of intrinsic properties such as gas content, metallicity, and star formation rate (§4) among dwarf galaxies. There are truly intermediate types which are probably in a transitional stage from Irr to dE (§7.6). A prototype dE/Irr in our neighborhood is the Phoenix system (van de Rydt et al. 1991). The Andromeda satellites NGC 205 and 185, too, are well-known "peculiar" dE's that contain dust and gas (e.g. Hodge 1971). Other, more distant examples of "mixed morphology" have been discussed by Sandage & Hoffman (1991) and Sandage & Fomalont (1993).

The following features are also relevant for the dE morphology (cf. Sandage & Binggeli 1984):

(1) *Nuclei.* Most bright dwarfs ($M_B < -16$) show a distinct luminosity spike in their center, commonly referred to as the central nucleus. These nuclei are not (yet) resolved at the distance of the Virgo cluster, i.e. they have a stellar appearance, but local resolved analogs, such as the nuclei of NGC 205 or M33, suggest that they are dynamically separate supermassive star clusters. The brightest nuclei can reach up to 20% of the total light of the parent dwarf galaxy (see Fig. 2) . The ratio of nucleated-to-normal dE's is monotonically decreasing with decreasing luminosity; faint dwarfs usually do not have a nucleus (Sandage et al. 1985b). The presence of a nucleus is conveniently indicated by appending an "N" to the type: dE,N.

(2) *Dwarf S0 types.* Sandage & Binggeli (1984) introduced the dwarf S0 type, which is a rare variation of the dE class. In their azimuthally averaged surface brightness distribution the dS0's are indistinguishable from bright dE' s (Binggeli & Cameron 1991). Some dS0's do show a pronounced two-component structure that is reminiscent of classical S0's, but there are a variety of other reasons why a dwarf was (and may be) called dS0 rather than dE (e.g. the presence of a bar feature, twisting isophotes, or simply high apparent flattening; see Binggeli & Cameron (1991). The dS0 class is very inhomogeneous and since it is so small (there are only about 25 dS0's in the Virgo cluster, as compared to 800 dE's) it will be mixed into the dE class, if not otherwise stated.

(3) *Huge, low-surface brightness types.* This is another class of galaxies first isolated in the Virgo cluster by Sandage & Binggeli (1984): very extended systems of extremely low surface brightness and almost no gradient, mostly classified as dE, but often also as intermediate (dE/Irr) or even irregular. These galaxies fall off the canonical dE sequence (see §2.2.2). The most extreme examples in the Virgo cluster have been found by Impey et al. (1988) using a special photographic technique developed by D. Malin. Similar galaxies exist in the Fornax cluster (Bothun et al. 1991). None are known in our local neighborhood.

Typical dwarf ellipticals, as well as some related types, are shown in Fig. 1.

Figure 1:
A collection of dwarf galaxy members of the Virgo cluster, adapted from Binggeli (1994a). The common scale is indicated on top. Typical dwarf ellipticals in the left row (c, f, i, l), along with the two "dS0" variants (a, d), are confronted with a low-luminosity, compact E (b), and with various types of smooth (g, h) or clumpy (k, m) dwarf irregulars. Two intermediate (transitional) types are also shown (e, n), the latter of which is a "huge, low-surface brightness" type. The individual names and types are as follows: a = NGC 4431 (dS0(5),N), b = NGC 4486B (E1), c = IC 3328 (dE1,N), d = IC 3435 (dS0(8),N), e = NGC 4344 (S pec, N:/BCD), f = IC 3457 (dE4,N), g = IC 3416 (ImIII), h = UGC 7636 (ImIII-IV), i = VCC 1661 (dE0,N), k = IC 3453 (ImIII/BCD), l = VCC 354 (dE0), m = VCC 1313 (BCD), n = IC 3475 (ImIV or dE2 pec).

## 2.2. Photometric Properties

Roughly 2000 dE's have been visually classified (from fine-scale photographic plates) on the Sandage & Binggeli (1984) system. Digitized photographic profiles have been measured for a small ($\sim 300$), not necessarily random, subset of these (Ichikawa et al. 1986; Binggeli et al. 1984; Binggeli & Cameron 1991). Published CCD photometry exists for a still smaller ($\sim 100$), still less random, subset (e.g. Caldwell & Bothun 1987; Impey et al. 1988; Bothun et al. 1991; Vader & Chaboyer 1994). Our knowledge of the quantitative photometric properties of these galaxies is thus still in its infancy.

The isophotes of dE's are typically well fit by ellipses. Analysis of the two-dimensional structure of the isphotes gives information on the intrinsic shapes of the galaxies (§2.2.3). The one-dimensional profiles (typically referred to the major axis, or to some mean axis) presumably also hold clues to how the galaxies collapsed and formed stars, and are the subject of the next section.

### 2.2.1. Surface-Brightness Profiles

The first dE profiles to appear in the literature were those of the six classical "dwarf spheroidals" derived by Hodge from star counts (see Hodge 1971 for detailed references). The pioneering work of Hodge has only recently been confirmed and superseded in accuracy by Irwin & Hatzidimitriou (1993). Hodge realized at once that neither Hubble's $r^{-2}$ law (Reynolds 1913; Hubble 1930), nor de Vaucouleurs' (1948) law ($I(r) = I_0 exp(-7.67(r/r_e)^{1/4})$ for giant ellipticals adequately describe the dwarf profiles. These looked almost box-like, i.e. flat in the inner part, followed by a sharp cut-off. A cut-off ("tidal") radius had earlier been introduced by King (1962) for star clusters, and this seemed to apply to the dwarfs as well. Perfect fits to the "spheroidal" profiles were finally achieved by Hodge & Michie (1969) by employing the multi-parameter model of Michie (1963), which also allowed a dynamical interpretation. However, in later times the simpler King (1966) model has taken over for the purpose of profile fitting (it is a special case of the Michie model, cf. Binney & Tremaine 1987).

Strangely (as it appears in retrospect), only in 1983 was it appreciated (Faber & Lin 1983) that the profiles of "dwarf spheroidals" are fairly good exponentials ($I(r) = I_0 \exp(-r/r_0)$, where $I_0$ is the central intensity and $r_0$ is the scale length) – just as those of irregulars and of the disk components of spirals and S0's are (e.g. de Vaucouleurs 1959; Freeman 1970; Carignan 1985; Bothun et al. 1986). Not all of them are nice exponentials (e.g., the Draco system), which must be the reason why the exponential was dismissed by Hodge, but such deviations are also common among irregular profiles. Better fits can always be achieved with 3-parameter forms like Oemler's (1976) modified Hubble law (Caldwell 1983; Binggeli et al. 1984), or King profiles (Binggeli et al. 1984; Ichikawa et al. 1986; Ichikawa 1989; Binggeli & Cameron 1991; Vader & Chaboyer 1994). However, the exponential is more economical with only 2 free parameters (central surface brightness and scale length) and it is almost universal for faint galaxies, allowing easy comparison of photometric parameters between different types of dwarfs. For a comparison with normal ellipticals, the King profile is still useful, because it applies to E's as well as dE's (see below).

Figure 2:
Surface brightness profiles of five dwarf ellipticals with prominent nuclei. The solid lines represent the sum of a King profile to the outer regions (with parameters $r_c, \mu_0$, and $\log(r_t/r_c)$), and a central point source (convolved with the appropriate PSF) fit to the inner regions such that the observed surface brightness is nowhere exceeded. The percentage of the total light contributed by the central light excess (nucleus) with respect to the King fit is given under the heading $f_{ex}$. From Vader & Chaboyer (1994).

Following Faber & Lin (1983), it was quickly shown that the exponential is also a good fitting law for Virgo cluster dwarf ellipticals (Binggeli et al. 1984), leading to the notion that there are *two separate families of elliptical-like galaxies*: diffuse dwarf ellipticals ("spheroidals") that have nearly exponential profiles *versus* classical ellipticals (including M32) that follow a de Vaucouleurs $r^{1/4}$ law (Wirth & Gallagher 1984). The structural similarity between dwarf irregulars and dwarf ellipticals also reopened the discussion, pioneered by Einasto et al. (1974), on a possible evolutionary Irr $\Rightarrow$ dE transition (Lin & Faber 1983; Kormendy 1985; Binggeli 1986; Bothun et al. 1985; Bothun et al. 1986; cf. §7.6 for the later development).

In spite of the current popularity of the exponential as a fitting law for dE profiles, it is important to note that only faint dE's ($M_B > -16$, the majority, to be sure) are good exponentials everywhere, i.e. over the *whole* radius range. Bright dE's ($M_B \leq -16$) usually deviate from an exponential in their inner part; there is a central surface brightness excess which, on average, is stronger in brighter galaxies (Caldwell & Bothun 1987; Binggeli & Cameron 1991). Binggeli & Cameron (1991) have introduced profile types for early-type galaxies based on the degree of linearity of the profile. Caldwell & Bothun (1987) identify the central light excess in a bright dE, i.e. the residual from the exponential fit to the outer part, as "the nucleus." However Binggeli & Cameron (1991) contend that there are two components to the central excess: one that is currently unresolved at Virgo distances, and the other extended over several arcsec. In most bright dwarfs one can clearly see such a sharp nucleus *on top of* a shallow, extended surface brightness excess. Faint dE's, on the other hand, frequently have nuclei, but rarely show the more extended excess. The distinctness and non-resolution (point-source nature) of some very bright dE nuclei is illustrated in Fig. 2, seen in this case on top of King, rather than exponential, profiles.

The profiles of the very brightest dE's and dS0's are almost as strongly curved as the profiles of normal ellipticals. Binggeli & Cameron (1991) attempt a separation between E's and dE's by going to the logarithmic (log $r$) representation of profiles, and by fitting King (1966) models to both E's and dE's. In the radius range 0.1 to 1 Kpc, the profiles of early-type galaxies appear to fall into two distinct classes: (1) the steep profiles of "normal" (or "classical") ellipticals and S0's, and (2) the significantly flatter profiles of "dwarf" ellipticals and S0's (called so *by definition*). The latter group includes all the "intermediate" (E/dE) cases, as judged from morphology. In terms of best-fitting King model profiles, the normal/dwarf ellipticals have high/low central surface brightnesses and small/large core radii, again supporting the notion of a *dichotomy* among elliptical stellar systems. However, the King model does not fit the bright dwarf profiles very well. As with the exponential fits, there remains a strong, extended surface brightness excess in the central part, which casts some doubt on the whole fitting procedure. When the central excess is ignored, the dichotomy between dE and E galaxies is pronounced. However, it is not clear that a strong dichotomy remains when model-independent central-surface brightnesses are used. Binggeli & Cameron (1991) were not able to resolve any of the low-luminosity (M32-type) ellipticals in their sample ("resolve" in the sense of Schweizer 1979; 1981 ); they could only surmise that these would be nearly as compact as M32. Prugniel et al. (1992) claimed resolution of the very same compact E's and reported a continuity of profiles between E's and dE's. However, Kormendy & Bender (1994), working with the superior resolving powers of CFHT and HST, make clear that the cores of those compact galaxies are yet to be resolved. It is thus still not clear whether E's and dE's form distinct sequences in their core parameters. In any case, the number of objects in this E/dE intermediate zone is very small (Prugniel 1994; Vader & Chaboyer 1994).

*2.2.2. Parameter Correlations and Selection Effects*

In photometric diagrams involving the model-independent *effective* parameters $r_e$ and $\langle\mu\rangle_e$, the mean surface brightness within $r_e$, early-type galaxies are confined to *one* broad sequence with a distinct, almost orthogonal break around $M_B \approx -20$, $\langle\mu\rangle_e \approx 21B$ mag per square arcsec, and log $r_e[kpc] \approx 0.3$ (Kodaira et al. 1983; Binggeli et al. 1984; Binggeli & Cameron 1991; Capaccioli et al. 1993). Galaxies on the bright branch have increasing surface brightness with decreasing luminosity (known since Kormendy 1977), while galaxies on the faint branch, i.e. the dwarfs, follow the inverse trend, with a scaling law of

$$\langle\mu\rangle_e \simeq 0.75 M_B + 35.3 \qquad (1)$$

for a Virgo cluster distance modulus of 31.7 (Binggeli & Cameron 1991). The scatter of this relation is considerable (0.8 mag at the 1 $\sigma$ level) but nevertheless allows the use of $\mu$ as a distance indicator (see §8.1). The identification of the two branches with two fundamentally different *sequences* of galaxies rests of course on the disparity, if not dichotomy, of the profiles discussed above.

In the King model representation of profiles, the dE sequence is essentially delineated by the central surface brightness, with $\mu_0 \propto M_B$. The core radius is roughly constant at $r_c \approx$ 1 kpc, albeit with much scatter. The third parameter, the concentration index log $r_t/r_c$, again tends to decrease with decreasing luminosity, i.e. the cut-off length of the dwarfs gets shorter and shorter (Binggeli et al. 1984; Ichikawa et al. 1986; Binggeli & Cameron 1991). In the context of King's (1966) dynamical model, $r_t$ has the meaning of a *tidal* radius. The observed $\log r_t/r_c - M$ relation is plausible, qualitatively, because low-mass galaxies will be more more easily and deeply stripped by tidal forces. The problem is to find a tidal agent. Binggeli et al. (1984) used the formulation of Richstone (1976) to calculate the cumulative tidal effect exerted on a Virgo cluster dwarf galaxy by the frequent encounters it has with other cluster galaxies. The effect turned out to be quite small. Another possibility was explored by Binggeli (unpublished) based on Merritt's (1984) model in which the size of cluster galaxies is determined by the tidal field of the cluster as a whole. The dwarfs could indeed just barely have been shaped by this process, which roughly also reproduces the slope of the observed $r_t - M$ relation. However, dE's outside clusters are, so far as we can tell, indistinguishable from cluster dE's. Also, faint dE's are perfect exponentials, just like the dwarf irregulars which cannot possibly be interpreted as King spheroidals. Thus the "tidal" radius is probably an *intrinsic* property of the dE's, and is probably not useful as a diagnostic of tidal agents (e.g. for probing the galactic halo as in Faber & Lin 1983).

Figure 3:
Schematic $M_B - \mu$ plane for stellar systems and subsystems. Dwarf ellipticals and dwarf irregulars, for $M_B > -16$, follow a common $M_B - \mu$ relation, which is roughly orthogonal to, and detached from (?), the $M_B - \mu$ relation of normal E's and bulges. For more details see text. Taken from Binggeli (1994b).

In the exponential representation of the profiles, the best-fitting exponential scale-lengths ($r_0$) and central surface brightnesses ($\mu_0$) of the dE's follow the same general trend; both are monotonically decreasing with decreasing luminosity. However, there is a break around $M_B = -16$ in the $r_0 - M$ and $\mu_0 - M$ relations (Binggeli & Cameron 1991): for bright dwarfs ($M_B < -16$) the relations are

$$\log r_0[\text{pc}] \simeq -0.2 M_B - 0.3\,, \quad \mu_0 \simeq 22.7 B/\square''\,; \qquad (2)$$

for faint dwarfs ($M_B > -16$) we have

$$\log r_0[\text{pc}] \simeq -0.02 M_B + 2.6\,, \quad \mu_0 \simeq 0.7 M_B + 34\,, \qquad (3)$$

again for a Virgo modulus of 31.7. The near constancy of $\mu_0$ of bright dwarfs is reminiscent of "Freeman's law" for spiral galaxies (Freeman 1970; van der Kruit 1987; see also Fig. 3). Binggeli & Cameron (1991) speculated that the inflection around $M_B = -16$ might reflect a transition from bright, rotation-supported stellar disks to faint, pressure-supported spheroidals. There are indeed many dS0's among the bright dwarfs, whose disk nature seems manifest. However, there are hints that bright dE's are non-rotating as well (Bender & Nieto 1990).

The location of dwarf ellipticals in the $\mu - M$ plane identifies them as a distinct class of object. Fig. 3, taken from Binggeli (1994b), shows a schematic $\mu - M$ diagram for all kinds of stellar systems and subsystems (disks, bulges, nuclei, star clusters). Here, $\mu$ is the model-independent, apparent *central* surface brightness in the sense of Kormendy (1985). The two sequences of dE's versus E's + bulges, and the problem of their connection (indicated by a question mark) have been discussed above. A third sequence of purely stellar systems, at low luminosities but high densities, is formed by globular clusters. Galaxian nuclei probably belong to the same sequence and extend it to higher densities and luminosities. The nuclei of M31 (Lauer et al. 1993) and M33 (Kormendy & McClure 1993), as well as the nuclear magnitudes of all dE,N's (Binggeli & Cameron 1991) fall indeed into the region indicated. A fourth sequence is defined by the disk components of spiral and S0 galaxies, and dwarf irregular galaxies. At the bright end, this sequence levels off at $\mu_0(B) \approx 21$ (Freeman 1970; van der Kruit 1987). Fainter than $M_B \approx -17$ or so, the central surface brightness starts to get fainter and fainter with decreasing luminosity, i.e. the sequence bends over to the $\mu - M$ relation of dwarf irregulars, which is identical to that of faint dE's, although clumpy irregulars ("BCD's" – not shown in the figure) tend to widen the distribution.

The close kinship of the dE and S+Irr sequences is of course suggestive of a common origin of the two classes of objects and an evolutionary connection in the direction S+Irr ⇒ dE (e.g., Kormendy 1985, 1986). However, *bright* dE's ($M_B < -16$) have systematically higher surface brightness than disk galaxies of the same luminosity, which is one of the reasons why a simple gas removal scenario does not work there (Bothun et al. 1986; Davies & Phillipps 1988; James 1991; Binggeli 1994a; cf. also §7.6). On the other hand, *faint* dE's and Irr's ($M_B > -16$) are often hard to distinguish with respect to morphology and even stellar content. There are many "intermediate" types of "mixed morphology" (cf. above). However, environments such as the Virgo cluster are overwhelmingly dominated (in number) by smooth-profile galaxies.

A separation into "dwarf" and non-dwarf ("normal," "giant," "classical") types *along* the S+Irr sequence is not as clear-cut as it appears between the E and dE sequences. A possible parallel definition to the one for early-type dwarfs has been suggested by Binggeli (1994a): a late-type galaxy should be called "dwarf" if it lacks an E-like bulge. This would include also many Sc spirals. The general definition for dwarf galaxies would then be: *dwarf galaxies have no E-component*. There are of course difficulties with such a definition (cf. Binggeli 1994a,b). In any case (but certainly for $M > -16$), one can speak of a common, broad sequence of dwarf galaxies, which is indeed the *main sequence of galaxies*.

There is a possible caveat. The photometric sequences in the $\mu - M$ plane might be shaped, if not produced, by selection effects. Both very compact, high-surface brightness objects and extended, very low-surface brightness objects have small isophotal diameters and can easily go undetected in a survey. A photometric sequence could just be what is *visible* within the detection limits. Reaves (1956) and Arp (1965) issued the first warnings to this effect. The issue was vigorously taken up and quantified by the Cardiff group led by M. Disney (Disney 1976; Disney 1980; Disney & Phillipps 1983; Disney & Phillipps 1987). A great boost for the suspicion that entire populations of galaxies are missing in our conventional catalogs was provided by the serendipitous discovery of "Malin 1," a very huge, very low-surface brightness spiral (Bothun et al. 1987). This object indeed falls off any sequence (see Fig. 3). Additional very low-surface brightness objects turned up in the Virgo (Impey et al. 1988) and Fornax (Bothun et al. 1991) clusters, obviously being cluster members. Earlier, Sandage & Binggeli (1984) had found many brighter examples of this class of "huge, low-surface brightness dwarfs" (cf. above). These objects progressively broaden the $\mu - M$ sequence of dwarf galaxies towards the faint end (see Fig. 3), posing severe problems for the determination of the faint end of the luminosity function (§5).

The existence of any $\mu - M$ relation for dwarfs was denied altogether by Davies et al. (1988) and Phillipps et al. (1988). For dE galaxies in clusters, the argument hinges on whether faint, relatively high surface-brightness objects are cluster members or background galaxies (Ferguson & Sandage 1988; Irwin et al. 1990a; Bothun et al. 1991). In deep surveys of the region of the Fornax cluster (Ferguson & Sandage 1988; Ferguson 1989; Bothun et al. 1991), the surface-density enhancement of the cluster is readily seen in the spatial distribution of classical dE galaxies (those that lie on the $\mu - M$ relation), while the cluster is not at all visible in the spatial distribution of galaxies that lie off the relation. However, Irwin et al. (1990a) see the number counts on the Fornax cluster UK Schmidt plate enhanced at all surface-brightnesses relative to a nearby comparison field. Furthermore, Phillipps et al. (1988) show that field dwarf irregulars may not follow an intrinsic surface-brightness luminosity relation. Overall, it must be admitted that there are several regions in the $\mu - M$ plane where we have a blind eye, i.e., where the visibility of objects is very low, and where more sequences of galaxian objects *might* one day be uncovered. However, the surveys for (in particular) low-surface brightness objects that have been undertaken so far suggest that the extremes of the surface-brightness distribution are not heavily populated. In the luminosity range well sampled by such surveys, the *ridge line* dE sequence is well defined and is certainly not an artifact of any detection limits, or visibility function (even if the full extent of the scatter is not yet well quantified). Hence the sequence must be a physical reality, and must have a physical origin, which is dealt with in §7.

*2.2.3. Flattenings*

Overall, dwarf ellipticals are slightly more flattened than normal E's (Binggeli & Popescu 1994; Ryden & Terndrup 1994). Previous studies suffered from strong systematic errors in the eye-estimated flattening data. Apparently round (ellipticity class E0) systems, among both E's (e.g., Binney & de Vaucouleurs 1981) and dE's (e.g., Sandage et al. 1985a), were grossly overabundant due to the tendency of the eye (mind) to see (force) perfect symmetry where there is only near symmetry. Reliable (photometrically measured) ellipticity distributions were determined for dE's (Ichikawa et al. 1986; Ichikawa 1989) even *before* the equivalent became available for E's (Franx et al. 1991; Ryden 1992).

There is good agreement about the apparent ellipticity distribution of dE's, except that Binggeli & Popescu (1994) find that only the non-nucleated dE's are significantly more flattened than the E's, while Ryden & Terndrup (1994) claim the same effect for nucleated dwarfs as well. That the dE,N's tend to be rounder than the dE(no N)'s has been noted before – without claiming significance – by van den Bergh (1986), Impey et al. (1988), Ferguson & Sandage (1989) and Ichikawa (1989). A dynamical explanation of the effect has been offered by Norman (1986), based on the work of Norman et al. (1985). These authors have shown that a dense central cusp in the center of a triaxial system with a substantial fraction of box orbits will, over many crossing times, make the galaxy rounder. For a nucleated dwarf, a decrease of one or two ellipticity classes over a Hubble time can be expected. However, whether the nuclei have been around for a Hubble time must be questioned. There are hints that some of the nuclei formed much more recently (Davies & Phillipps 1988; Meurer et al. 1992). Alternatively, the more flattened dwarf systems, due to higher (?) angular momentum, might have avoided substantial infall of gas into their centers, thus preventing the formation of a nucleus. The problem here is that flattening and angular momentum are not clearly correlated: neither bright nor faint dE's appear to be rotation-supported (Bender & Nieto 1990; Bender et al. 1992).

The surface brightness test for the intrinsic shape of elliptical systems (Marchant & Olson 1979; Richstone 1979), if taken at face value, clearly favours the oblate over the prolate spheroid – for normal ellipticals (e.g., Fasano 1991, and references therein) and dwarf ellipticals alike (Ichikawa 1989; Binggeli & Popescu 1994). Both E's and dE's are more likely triaxial in shape. The pure oblate model is barely able to reproduce the paucity of apparently round E's (Franx et al. 1991) or dE's (Binggeli & Popescu 1994).

The flattening distributions of smooth dwarf irregulars (Im's) and of (at least the) non-nucleated dE's are surprisingly similar (Binggeli & Popescu 1994), confirming earlier findings by Ichikawa et al. (1986), Feitzinger & Galinski (1986) and Ferguson & Sandage (1989), however disproving the often cited result of Sandage et al. (1985a). This supports the view that the non-nucleated dE's are the dead remnants of formerly star-forming dwarf galaxies that, by internal or external causes, lost or consumed their gas (Davies & Phillipps 1988; Thuan 1992; Binggeli 1993). However, the fact that the *nucleated* dE's are rounder than the late-type dwarfs does not, per se, exclude the possibility that they, too, are such remnants. There are various ways for a star-forming disk to puff up along its – presumably violent – transition into a purely stellar system (Biermann & Shapiro 1979; Farouki & Shapiro 1980; Davies & Phillipps 1988). Therefore, the shapes of dwarf galaxies, and of galaxies in general, do not provide very strong constraints on evolutionary connections between the different types (cf. also §7.6).

## 3. Kinematics and Dynamics

### 3.1. The Fundamental Plane

Giant elliptical galaxies occupy only a small portion of the three-dimensional parameter space defined by the central velocity dispersion ($\sigma_0$), effective surface brightness ($I_e$) and effective radius ($r_e$). The manifold occupied by ellipticals is most simply represented as a scaling law

$$r_e \propto \sigma_0^A I_e^B \qquad (4)$$

with $A \approx 1.4$ and $B \approx -0.9$ (Kormendy & Djorgovski 1989; Bender et al. 1992). Simple arguments from the virial theorem and the above scaling relation suggest that $M/L \propto M^{1/6} \propto L^{1/5}$ (Dressler et al. 1987). Properties of the stellar populations appear to be closely related to structure and dynamics; tight relations also exist between $r_e$, $I_e$ and color or metallicity (de Carvalho & Djorgovski 1989). While the existence of a "fundamental plane" for elliptical galaxies provides important clues about the formation of these galaxies, the implications of these clues are currently understood only at a qualitative level (Kormendy

& Djorgovski 1989; Bender et al. 1992).

Several attempts have been made to assess whether dE galaxies lie on the fundamental plane defined by the giants. The task is complicated by the great difficulty in measuring velocity dispersions for faint, low-surface-brightness galaxies. The faint end of the luminosity function is represented by Local Group companions, while the brighter dE's for which velocity dispersions have been measured are mostly members of the Virgo cluster. These latter samples are far from complete, and are biased toward high-surface-brightness objects (Bender & Nieto 1990; Peterson & Caldwell 1993). Furthermore, the prominent nuclei in many of the brighter dE's may be distinct dynamical entities, so using *central* velocity dispersions may not be sensible if correlations of global parameters are sought.

Nieto et al. (1990) presented the first attempt to extend the fundamental plane to low galactic mass. Their analysis suggested that dE's and also globular clusters fall near or within the fundamental plane, albeit with more scatter than expected from the measurement errors. More recent analysis suggests that dE's do not follow the giant-elliptical scaling relations (Bender et al. 1992; de Carvalho & Djorgovski 1992; Peterson & Caldwell 1993). Fig. 4 shows the distribution of low-luminosity ellipticals (not all of the diffuse kind) along one projection of the fundamental plane. The local dE's evidently depart from the giant E fundamental plane. The two sequences appear to intersect at $M_B \approx -17$, in the regime of the brighter cluster dE's. However, the situation is rendered ambiguous by the large scatter in the dE measurements, and by an apparent inconsistency between the two largest datasets. Bender & Nieto (1990) measured $\sigma$ for seven low-luminosity ellipticals. However their sample was manifestly biased toward high-surface-brightness galaxies, and interacting ones at that. Only two of their galaxies fall on the canonical dE sequence, and these had central velocity dispersions of $65\pm4$ and $52\pm6\,\mathrm{km\,s^{-1}}$. Peterson & Caldwell (1993) measured eight additional dE's, all but one of the nucleated variety. They found $16.4 < \sigma < 39\,\mathrm{km\,s^{-1}}$. It is not clear whether the discrepancy with Bender & Nieto (1990) is due to the different types of galaxies in the samples, the different spectral resolutions, or different apertures. It is clear that measurements are needed for a *complete* sample that is unbiased in surface-brightness to determine the true relations for dE's between structure and velocity dispersion. Measurements for cluster dE's with absolute magnitudes fainter than -15, while exceedingly difficult to obtain, are important for testing whether cluster dE's and local dE's really are the same type of object.

Dwarf ellipticals appear to define more nearly a one-parameter family, when $L, r_e, I_e$, and $\sigma_0$ are considered; the scatter in the surface-brightness–magnitude relation (§2.2.2) is not significantly reduced if surface-brightness is replaced by a linear combination of surface-brightness and $\log \sigma_0$. However, the paucity of velocity dispersions for galaxies between $M_B = -16$ and $M_B = -11$ renders this to a certain extent an exercise in small-number statistics. The statistics can be improved by substituting color for velocity dispersion. For Virgo and Fornax cluster dE's, there is a small improvement if color is added as an additional parameter, but large scatter remains. The rather heterogeneous samples of galaxies for which colors have been measured seriously compromises the analysis of scaling relations and tests for the number of independent parameters. While giant E samples are reasonably free from selection effects based on color and surface-brightness, dE samples are greatly affected by such selection effects. Low surface-brightness galaxies are typically detected on blue-sensitive emulsions; hence the discovery technique will tend to find blue galaxies and miss red ones if there is a spread in color at fixed bolometric surface-brightness. Furthermore, photometry samples in the literature are a mix of bright, relatively high-surface brightness dE's chosen for ease in getting photometry (Bothun & Caldwell 1984; Caldwell & Bothun 1987), and extremely low surface-brightness dE's chosen because they are interesting (Impey et al. 1988; Bothun et al. 1991).

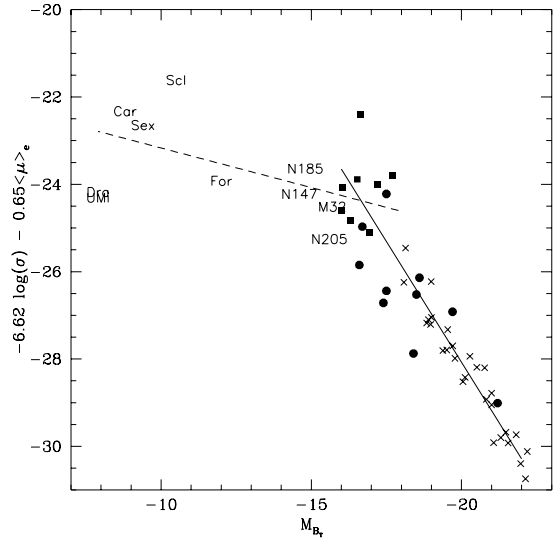

Figure 4:
One projection of the fundamental plane of elliptical galaxies. Data from Dressler *et al.* (1987) for Virgo and Fornax cluster ellipticals are shown as x's. Galaxies surveyed by Bender & Nieto (1990) are shown as circles, while those of Peterson & Caldwell (1993) are squares. Local group galaxies are shown by their abbreviations. Data for the Local Group galaxies comes from the compilation by Bender *et al.* (1992), with the exception of Sextans, for which we have used the values in Peterson & Caldwell (1993). The solid line is the least squares fit to the Dressler *et al.* Virgo-cluster data. The dotted line shows the trend for $M/L \propto L^{-0.37}$ predicted by Dekel & Silk (1986). At fixed luminosity, the velocity dispersions reported by Peterson & Caldwell are uniformly lower than those reported by Bender & Neito, rendering it difficult to tell which sequence the cluster dE's actually inhabit.

The analysis of the existing samples suggests correlations that are in reasonable agreement with the predictions of Dekel and Silk (1986). Specifically, while the fundamental-plane relation for giant ellipticals implies $M/L \propto L^{1/5}$, the relation for dwarfs is $M/L \propto L^{-0.4}$, close to the $M/L \propto L^{-0.37}$ relation expected for mass-loss within a dominant dark halo (Dekel & Silk 1986). However, it must be noted that the observed relation has large scatter (and involves many assumptions) and there is a hint of a correlation of $M/L$ for local dE's with position in the outer-galaxy halo (see below).

### 3.2. (Non–) Rotational Support

The question of rotation in dwarf ellipticals has been addressed only recently. Obviously, the measurement of rotation requires spectroscopy at an even fainter level of surface brightness than measurement of a central $\sigma$. The sparse data we have to date suggest that *dwarf ellipticals are not supported by rotation.* This excludes at least some "dS0" systems, which we have generally lumped together with the dE class. Data are published for two bright dE's in the Virgo cluster, VCC 351 and IC 794 (Bender & Nieto 1990); the three bright dE companions of M31: NGC 205 (Carter & Sadler 1990; Held et al. 1990; Bender et al. 1991), NGC 185 (Bender et al. 1991; Held et al. 1992), and NGC 147 (Bender et al. 1991); and the resolved Fornax system (Paltoglou & Freeman 1987; Mateo et al. 1991). A convenient measure of the dynamical importance of rotation is provided by the anisotropy parameter

$$(v/\bar{\sigma})^{\star} = \frac{v/\bar{\sigma}}{\sqrt{\epsilon/(1-\epsilon)}} \qquad (5)$$

where $v$ is the rotational velocity $\bar{\sigma}$ is the mean velocity dispersion, and $\epsilon$ is the ellipticity at the radius where $v$ is measured (Binney 1978; Kormendy 1982) . If $(v/\bar{\sigma})^{\star} \geq 1$, the object can be assumed to be flattened by rotation; if $(v/\bar{\sigma})^{\star}$ is significantly smaller than 1, the object must be supported by velocity anisotropy. The six dwarfs from above all have $(v/\bar{\sigma})^{\star} < 0.4$, hence they are clearly not rotation-supported. As mentioned, the dS0's may not follow this trend, e.g., UGC 7436, a dS0(5),N system in the Virgo cluster is a rotating disk galaxy (Bender, private communication). Many dS0's exhibit disk-like morphologies (Binggeli & Cameron 1991, cf. §2), but their kinematic nature has yet to be explored.

Bender & Nieto (1990) have also measured a number of faint classical and compact E's in the Virgo cluster, which all seem to be rotation-supported, in accord with the trend found by Davies et al. (1983). An exception to this rule is the very compact galaxy M32 with $(v/\bar{\sigma})^{\star} \approx 0.5$ (Tonry 1984; Tonry 1987; Dressler & Richstone 1988), but M32 is likely tidally influenced by M31. The separation between *rotating low-luminosity E's* and *anisotropic dE's* apparently goes hand in hand with the photometric discontinuity between E's and dE's described above (§2.2). However, a much larger kinematic sample is needed to test for a real discontinuity in $(v/\bar{\sigma})^{\star}$. Bender & Nieto (1990) argue that the *specific angular momentum*$(J/M)$ of the dwarfs, unlike that of giant ellipticals, falls onto the $J/M - M$ relation for rotating ellipticals. The dwarfs seem to compensate for their slow rotation with a larger effective radius. Bender & Nieto (1990) suggest that dE's gained their anisotropy from expansion due to winds after the onset of star formation, but admit that gas accretion and tidal interactions could be important as well. Shaya & Tully (1984) have suggested that the low angular momentum content of elliptical galaxies is due to tidal interaction with the collapsing protocluster. Such a mechanism may also apply to dE galaxies, although their higher specific angular momentum may be somewhat problematical.

### 3.3. Local dE's and Dark Matter

Much of the recent upsurge in dE-related research is due to the problem of dark matter (hereafter DM). With decreasing luminosity, galaxies appear to become increasingly dominated by DM (e.g., Kormendy 1988). The dE ("dwarf spheroidal") companions of our Galaxy, which include the intrinsically faintest galaxies known, have thus become prime targets for the study of DM. The first indication of DM in the local dwarfs is due to Aaronson (1983). His bold announcement of an unusually high velocity dispersion in Draco (based on only three stars) has essentially stood the test of time. In the following we give a brief account of the present status of the search for DM in the local dE's, drawing on the excellent reviews of Mateo (1994), Pryor (1994), and Gerhard (1994) delivered at the recent ESO/OHP workshop on "Dwarf Galaxies" (but see also Pryor 1992, and Mateo et al. 1993).

The observational task is formidable: one has to measure precise radial velocities (with errors smaller than 10 km s$^{-1}$) for as many as possible, apparently faint (V > 17.5) and weak-lined (low-metallicity) stars. Nevertheless, measurements have now been made for all but one (Leo I) of the eight dE companions of the Galaxy, with varying quantity (number of stars) and quality of data. With the exception of the best-studied Fornax system (Paltoglou & Freeman 1987), the only kinematic datum available is the central $\sigma$, which is typically 10 km s$^{-1}$. There is little room left for doubts about the reality of the stellar movements. The only severe problem is contamination with binary stars, which can in principle be solved by finding the binaries through repeat measurements. Multi-epoch observations for some of the dwarfs suggest a binary frequency for giant stars of 10 - 15 %, which has little (though not negligible) effect on $\sigma$ (Mateo 1994).

Under the most simple assumptions that (1) mass follows light, and (2) the velocity distribution is isotropic, the calculation of a central mass density, $\rho_0$, and a mass-to-light ratio, $M/L$, is straightforward by "King's method" (King & Minkowski 1972), also called "core-fitting" (Richstone & Tremaine 1986). Present typical uncertainties in the photometric parameters ($\mu_0$ and $r_c$) and distances of the dwarfs, in addition to a 20 % uncertainty in $\sigma$, introduce errors of up to 50 % in $\rho_0$ and $M/L$. The derived $\rho_0$ values are generally in the range 0.1 - 1 M$_\odot$ pc$^{-3}$, while $M/L$ ranges from $\approx$ 5 (Fornax) to > 100 for Draco and Ursa Minor. As $M/L \approx 2$ for globular clusters, the presence of large amounts of dark matter is clearly indicated. There is a strong correlation betwen $M/L$ and total luminosity, as shown in Fig. 4 (left panel): the less luminous a dwarf the higher its $M/L$. This is the above-mentioned fundamental-plane relation for dwarf ellipticals, $M/L \propto L^{-0.4}$. Interestingly, the *total* masses of the dwarfs derived under these simplified assumptions is always a few times 10$^7$ M$_\odot$, which Mateo et al. (1993) suggest is a *minimal* mass for dark halos. From the start (Aaronson 1983; Lin & Faber 1983), the dark halos of the dwarfs were used to put constraints on the neutrino mass, and essentially to exclude neutrinos as DM constituents (Gerhard & Spergel 1992a).

However, there is no reason (beyond simplicity) to assume that mass follows light in the dwarfs; spiral galaxies clearly suggest that the dark halos are much more extended than the visible parts. Second, the lack of rotation (see above) manifestly shows that the assumption of velocity isotropy cannot be correct. Are there any constraints on the potential and distribution function of the dwarfs? It has been shown that not even *perfect* knowledge of the surface brightness and velocity dispersion profiles of a stellar system can uniquely determine its potential (Binney & Mamon 1982; Merritt 1987; Dejonghe & Merritt 1992). Only the most sophisticated use of the entire information contained in the positions and velocities of $\approx$ 1000 stars per galaxy might remove that indeterminacy in the future (Pryor 1994). At present lower limits on $\rho_0$ and $M/L$ can be obtained by exploring the allowed parameter space with 2-component models (Pryor & Kormendy 1990; Lake 1990), and by applying the virial theorem (Merritt 1987). The minimal $\rho_0$ from the virial theorem corresponds to constant mass density throughout the visible galaxy and to orbits that are strongly radial. The resulting $\rho_0^{\min}$ is about 10 times smaller than $\rho_0$ from King's method. But a galaxy with $\rho_0 = \rho_0^{\min}$ has a higher global $M/L$ than a galaxy with higher $\rho_0$. The minimal total mass required by the virial theorem corresponds to a point mass (black hole) at the center (which cannot be excluded at present, see Pryor 1994). The resulting global $(M/L)^{\min}$ is down by a factor of $\approx 3$ as compared to $M/L$ derived from King's method. Thus, there is apparently no way to avoid the need for DM in the local dE's (excepting perhaps Sculptor and Fornax). Pryor (1994) concludes that Draco and Ursa Minor, the two most extreme cases, must have central densities larger than $\approx 0.1$ $M_\odot$ pc$^{-3}$ and global $M/L$'s larger than $\approx 30$.

Figure 5:
Inferred mass-to-light ratio of the local dwarf elliptical galaxies (Milky Way companions) against luminosity (left panel), and against galactocentric distance (right panel). Taken from Gerhard (1994), based on data compiled by Mateo (1994).

Alternatives to a high DM content call into question the assumption of virial equilibrium, or, at a more fundamental level, the validity of Newtonian gravity. This latter possibility, while it can never be strictly excluded, has become less viable since Gerhard & Spergel (1992b) demonstrated that Milgrom's (1983) Modified Newtonian Dynamics (MOND) does not work for the local dwarfs. However, the first alternative, i.e. that the dwarfs are not in equilibrium but are in a state of tidal dissolution, has proven more persistent. Consider Fig. 5 (right panel). With the exception of Leo II (whose high $\sigma$ has yet to be confirmed), the dwarfs show a nice correlation between $M/L$ and galactocentric distance: the closer a dwarf the higher its $M/L$. The possibility that the closer dwarfs have been tidally heated by the Galaxy, thus *mimicking* a high $M/L$, was first explored by Kuhn & Miller (1989). The problem with this scenario is that the time scale of disintegration, in the absence of a dark halo, is rather short ($\approx 10^8$ years), which would make our coexistence with this event very coincidental. Kuhn's (1993) attempt to stretch the survival time of the dwarfs seems unconvincing (Gerhard 1994). Moreover, global tides induce large streaming motions rather than large $\sigma$'s (Piatek & Pryor 1994). On the other hand, dense halo objects (MACHO's) as tidal agents would require unrealistically high perturber masses, while the more promising mechanism of tides from "halo lumps" has yet to be explored (Gerhard 1994).

On the observational side, there are hints that tidal effects might be important. Faint extensions have been reported near Ursa Minor (Irwin & Hatzidimitriou 1993) and Sextans (Gould et al. 1992). Gerhard (1994) finds clear signs of streaming motion in Sextans. However, a tidal origin for the high apparent $M/L$ of Draco, Carina, and Ursa Minor appears unlikely.

## 4. Stellar Content

Understanding the star-forming histories of dwarf galaxies is of vital importance if we are to make the connection to cosmological issues such as the nature of faint blue galaxies or the shape of the halo mass function. Making such a connection requires that we understand the *statistical* properties of dwarf galaxies in general. Examples of peculiar star-forming histories in galaxies such as NGC 205 are a clear warning that we do not really know what regulates star formation in dE galaxies, but are of no use for larger issues until we know *how prevalent* such phenomena are. Do cluster dE's share the same episodic star formation seen in local dwarfs (§4.1)? Do they follow the same scaling of metallicity with luminosity, surface brightness, and velocity dispersion?

In models where star formation in dE's is governed purely by internal processes (e.g. feedback from supernovae or OB stars), the stellar populations of dE galaxies should be purely determined by their structural parameters, and not by their environment. However, there are a number of observations that suggest that environment is important in governing both the numbers and the morphology of dwarf ellipticals. Examples are: (1) evidence of multiple episodes of star formation in local dE's, reviewed below; (2) confinement of most Milky Way companion dE's to the plane of the Magellanic stream (§6.2); (3) systematic variations in the dwarf/giant ratio from rich clusters to loose groups (4) significant differences in the spatial distribution of nucleated and non-nucleated dE's in the Virgo and Fornax clusters, the nucleated dE's being more centrally concentrated (see §6); and (5) large scatter in the trends of color and absorption-line strengths (Brodie & Huchra 1991) with luminosity among cluster dE's (see §4.2).

### 4.1. The Local Dwarfs

The Local Group (LG) dE's provide the most direct evidence for that something more complicated must be at work than simple expulsion of gas after the cumulative energy input to the gas from star formation exceeds the binding energy. As there have been several thorough reviews of stellar populations in Local Group galaxies in the last few years (Hodge 1989; da Costa 1991,1994; van den Bergh 1994), we provide only a very brief summary. The existence of an ancient, globular cluster age, population in dE galaxies is inferred primarily from the presence of RR-Lyrae stars. With the possible exception of Leo I, all LG dE's have at least a small population of $\sim 15$ Gyr old stars. The existence of an intermediate, 2-10 Gyr age, population is inferred from the presence of carbon stars on the AGB with luminosities above that of the first giant branch tip, and, in some cases, by the presence of main-sequence turnoff stars with luminosities exceeding the turnoff for a globular cluster age population.

The mix of ages varies widely from galaxy to galaxy. Ursa Minor consists almost entirely of old stars, while Carina shows evidence of two widely separated star formation epochs, with the majority of the stars having formed in the second episode, $\sim 6$ Gyr ago (Mighell 1990; Smecker-Hane et al. 1994). The star-formation histories inferred from color-magnitude diagrams are summarized schematically in Fig. 6. There is some evidence that the nearer companions to the Milky Way and Andromeda galaxies have the largest proportion of old stars (van den Bergh 1994).

Figure 6:
Schematic plots of star-formation rate versus time, spanning the past 15 billion years, for local dwarf galaxies. Individual panels are arranged in order of decreasing galactocentric distance $R$ (in kpc). The figure shows that dwarfs close to the Galaxy formed most of their stars long ago, while more distant dwarfs contain young or intermediate-age stars. From van den Bergh (1994).

The star formation histories of local dE's are clearly very complex. Evidently several of the galaxies formed some stars, then waited a few billion years to form the rest. The reasons for this are not at all understood. With a refurbished HST, we may expect much deeper color–magnitude diagrams for the Local Group dwarfs. Reaching below the main-sequence turnoff, these diagrams should yield much more detailed star-formation histories, perhaps giving us some clues to the underlying physics.

### 4.2. Colors and Spectra of Cluster dE's

Given the strong evidence for intermediate-age and even young stars in Local Group dE's, it is worth asking whether cluster dE's show similar phenomena. Detecting intermediate age populations at the modest levels seen in most local dwarfs is virtually impossible without resolving the stellar population. However, there are numerous examples of dE-like galaxies with either slightly disturbed morphologies (Sandage & Binggeli 1984; Sandage & Hoffman 1991; Sandage & Fomalont 1993), unusual colors (Vigroux et al. 1985), or Balmer absorption features in their spectra (Bothun & Mould 1988; Gregg 1991). Clearly not all cluster dE galaxies have been inert for the last 10 Gyr.

In general, however, dE galaxies fall roughly on the extrapolation of the color–luminosity relation for giant ellipticals (Caldwell 1983; Caldwell & Bothun 1987; Prugniel et al. 1993). The colors of faint dE's are closer to those of globular clusters than giant ellipticals. Indeed, comparison to galactic globular clusters provides a possible test of whether the blue colors are entirely due to low metallicity, or whether youth also plays a role. To try to disentangle age and metallicity Thuan (1985) and Bothun et al. (1986) observed a subset of these dwarfs in the infrared $J, H$ and $K$ bands. Thuan concluded that the Virgo dE's have metallicities $Z_\odot/3 \leq Z \leq Z_\odot$ and ages since the last burst of star formation of 1 to 8 Gyr. However, the conclusion that dE's typically have an intermediate-age population is based on a rather dated set of models computed for stellar populations of solar metallicity (Struck-Marcell & Tinsley 1978). Bothun et al. (1985) noted that while dE's for the most part fall on the globular-cluster locus in the $B - H$ vs. $J - K$ plane, a few have bluer $B - H$ colors than globular clusters with the same $J - K$, suggesting younger ages.

Ferguson (1994) compiled line strengths and colors for the rather heterogeneous sample available in the literature and compared dE's, giant E's, and globular clusters to two recent population synthesis models (Worthey 1992; Bressan et al. 1994). The zeroth-order question addressed was whether *the majority* of stars in dE's could be younger than those in globular clusters. While dE's and globular clusters show significant differences in several of the line-strength–color relations plotted (cf. Ferguson 1994 Fig. 2), the differences are not easily attributable to age; they are more likely due to calibration problems, selection effects, or deficiencies in the population-synthesis models used for comparison. In most diagrams, the dE's follow the globular cluster locus, albeit with much scatter.

Held & Mould (1994) obtained spectra for 10 nucleated dE's in the Fornax cluster and examined both kinematics and stellar populations. Their most robust result is that dE colors (UBV) are tightly correlated with metallicities derived from line strengths. For galaxies in a narrow range of luminosities, they see metallicities that vary over 0.7 dex, with no clear correlation with either luminosity or surface brightness. They echo the findings of Bothun & Mould (1988) that (nucleated) dE's show a larger range of Balmer-line strengths at fixed metallicity than globular clusters. Galaxies with strong Balmer absorption are thought to have intermediate age populations. However, the dE's basically follow the globular-cluster locus in the $H\beta$ vs. [Fe/H] plane, and, interestingly, both Bothun & Mould (1988) and Held & Mould (1994) find a few dE's with Balmer lines significantly *weaker* than globular clusters of the same metallicity. Possible explanations include different initial mass functions or horizontal branch morphologies for those galaxies. As the metallicities overlap the range where the "second parameter effect" operates in globular clusters (see Fusi Pecci et al. 1993 for a recent discussion), variations in HB morphology are not inconceivable and could in principle account for both strong and weak Balmer lines without requiring intermediate-age stars.

In summary, only a few cluster dE's (about 10%) show *significant* evidence for a young or intermediate-age population from their spectra or colors. Most of the dE's could have ages of the bulk of their population ranging anywhere from about 5 Gyr to the age of globular clusters.

### 4.3. Correlations Between Stellar Populations and Structure

Correlations between stellar populations and structure provide some of the most important constraints on models for galaxy formation. The color–luminosity relation for elliptical galaxies has been known for more than three decades (Baum 1959; de Vaucouleurs 1961). A similar correlation exists for absorption-line strengths (Faber 1973), and has been interpreted as a metallicity–mass relation that is a natural consequence of galactic winds. Among giant ellipticals, metallicity indicators correlate more tightly with velocity dispersion than with luminosity (Terlevich et al. 1981; Burstein et al. 1988; Bower et al. 1992; Bender et al. 1993). Compact galaxies such as M32 and NGC 4486B have values of $Mg_2$ that are high for their luminosities, but entirely in keeping with their velocity dispersions.

Bender et al. (1993) examined the $Mg_2$–$\sigma$ correlation for a large sample of galaxies that included dE's from the Local Group and the Virgo cluster. They found that the same correlation extends from giant ellipticals to the faintest local dwarf spheroidals, although that statement is somewhat dependent on the adopted metallicity–$Mg_2$ conversion and contradicts the earlier conclusion of Vader (1986). Residuals from the mean trend are most likely due to young or intermediate-age stars in a small fraction of the galaxies. Schweizer et al. (1990) showed that among the giants the residuals from the mean $Mg_2$–$\sigma$ relation correlate with the "fine structure" parameter, a measure of the degree of morphological peculiarity. The data for the dwarfs are as yet too sparse to make the same test, but there are certainly examples of dE's with morphological peculiarities and evidence of young stars (e.g. NGC 205).

For giant ellipticals, the overall scatter in the Bender et al. (1993) Mg$_2$–$\sigma$ relation is only slightly larger than that expected from observational uncertainties. This small scatter, and the lack of any detectable correlations with other parameters (ellipticity, isophotal shape, or velocity anisotropy) is of fundamental importance. For example, the small scatter in the color–luminosity relation indicates that giant ellipticals must either be very old ($z_{\text{formation}} \gtrsim 2$), or have formed in a coordinated event at lower redshift (Bower et al. 1992). For giant E's the rms dispersion in age or metallicity at fixed $\sigma$ is less than about 15%. The rather limited data for dE galaxies allows variations of about 50%. Differences between nucleated and non-nucleated dE's have been sought (Caldwell & Bothun 1987; Evans et al. 1990), but not in a way that controls for the color–luminosity or color–velocity dispersion relations. Comparison of the colors of the nuclei and the envelopes show no systematic differences; however examples of both blue and red nuclei are known (Chaboyer 1994).

Dwarf ellipticals, with mean surface-densities of stars comparable to those of the *outer parts* of giant ellipticals, are useful for testing ideas for the origin of color and metallicity gradients. Such gradients (de Vaucouleurs 1961; Thomsen & Baum 1987; Gorgas et al. 1990) are not explained by the simplest models of galactic winds, which are typically one-zone models with a single criterion for ejection of gas from the galaxy (Arimoto & Yoshii 1987; Dekel & Silk 1986; Bressan et al. 1994), but can be produced naturally when variations in the starformation feedback efficiency with radius are taken into account. Franx & Illingworth (1990) suggested that the local metallicity within ellipticals is determined by local escape velocity, consistent with models involving an extended period of gaseous infall, or localized ejection of the ISM by supernovae. They base their argument on the color profiles for 17 ellipticals that follow roughly the same relation in a color *vs.* $v_{\text{esc}}$ plot. However, the small data set, the lack of a detailed comparison of $v_{\text{esc}}$ *vs.* other parameters such as surface-brightness or $r/r_e$ in reducing the scatter among the color gradients, and the rather large zeropoint uncertainties in the Franx & Illingworth (1990) dataset allow room for other possibilities. Bender et al. (1993) suggest that stellar populations $P$ are controlled by a combination of *total* mass $M$ and the local stellar volume density $\rho$ as $P = f(M^\alpha \rho^\beta)$, translating for observed quantities into Mg$_2 \propto 0.33 \log(M^2 <\rho>)$ and $B - V \propto 0.037 \log(M^2 <\rho>)$.

That such relations must be missing part of the story can be seen by plotting local colors *vs.* local line strengths (Gonzalez 1993; Ferguson 1994). If stellar populations were determined entirely by galaxy structure, the different measures of age and metallicity should track each other closely. In actuality, the slopes differ from galaxy to galaxy, do not follow the trend expected for pure metallicity variation, and do not extrapolate to the colors and line-strengths observed for dE galaxies. It is not yet clear whether this is a second-order effect (e.g. due to the fact that the *mix* of metallicities must vary as a function of radius in galaxies due to projection), or whether something more fundamental is at work. The key question is whether the stellar populations of dE's resemble those in the outer regions of giant ellipticals, and whether the resemblance is closer if the comparison is made at constant surface brightness or constant velocity dispersion.

More precise measurements of dE stellar populations are desperately needed. The well-observed local dwarfs give us some hint of a universal metallicity–luminosity relation, but also strong indications that the simplest models for explaining such a relation (supernova winds during a single rapid star-formation episode) must be wrong. Available data on cluster dE's neither strongly support nor rule out a similar relation between luminosity and metallicity. Their similar morphologies suggest that cluster and Local Group dE's share a common origin, but the much higher dwarf/giant ratio and the spatial segregation of different dE types within clusters (see §6) argues otherwise. For cluster dE's, the key test will be to determine, for a large homogeneous data set, which of the many possible parameters (velocity dispersion, surface brightness, luminosity, position in a cluster, presence or absence of a nucleus, etc.) influence the global stellar populations.

## 5. Luminosity Functions

The luminosity function (LF) of galaxies reflects both the initial conditions (e.g. the power spectrum of density fluctuations in the gravitational instability scenario) and the complicated physics of collapse, cooling, star formation, and feedback that govern how mass is converted into light. Based on arguments for hierarchical structure formation (Press & Schechter 1974), Schechter (1976) proposed the following functional form for the luminosity function:

$$\phi(L)dL = \phi^*(L/L^*)^\alpha e^{-L/L^*} d(L/L^*), \tag{6}$$

where is $\phi(L)$ is the number of galaxies per unit volume per unit luminosity, $L^*$ is a "characteristic" luminosity, and $\alpha$ is a "characteristic" faint-end slope. The arguments behind the Press & Schechter (1974) formalism have recently been made more rigorous (Bond et al. 1990; Bower 1991), but these apply to the collapse of galaxy halos, and do not directly translate into a luminosity function. The various attempts to incorporate star-formation physics into hierarchical models (§7) produce a break at $L^*$, due to the requirement that galaxies radiatively cool while they are collapsing. The power-law form at faint magnitudes is imposed by the initial power spectrum but modified by a mass-dependent star-formation efficiency due to galactic winds and by galaxy mergers.

While the Schechter function is a good first-order approximation to the overall LF, there is information as well in the *type specific* luminosity functions, which are typically not well-fit by Schechter functions. Binggeli et al. (1988) reviewed the type-specific LF in clusters and the field and hypothesized that the LF's of the individual morphological types are constant, independent of environment. The observed environmental variation in the total LF (summed over all types) could then be explained as a simple change in the morphological mix with environment, via the morphology–density relation (see also Thompson & Gregory 1980; Jerjen et al. 1992). The theoretical implication is that the physics that governs the LF within a given Hubble class is (to first order at least) independent of environment, while the creation/destruction processes that govern the mix of types are not.

The faint end of the LF is dominated by dE galaxies in clusters, while Sd and Im galaxies dominate in lower-density environments. A working hypothesis (Binggeli et al. 1988) is that variations in the faint end slope of the LF can be explained *entirely* by variations in the relative proportions of dE and Sd-Im galaxies. However, before assessing this hypothesis, we must highlight the uncertainties in current LF estimates and the selection effects that could bias the faint-end slope.

## 5.1. Selection Effects

One of the major problems in comparing models to the observed luminosity function (LF) is that galaxy surveys are limited by the brightness of the night sky and by the desire to distinguish stars from galaxies. This imposes limits on the detectability of both high and low surface-brightness galaxies (§2.2.2). The standard assertion that a survey is "magnitude limited" assumes (1) that the isophotal threshold of the survey encompasses most of the light of *all* types of galaxies, and (2) that star–galaxy separation is not a problem. Neither assumption is safe for dwarf galaxies.

Figure 7 illustrates the importance of the isophotal threshold. Shown at left are surface-brightness profiles for three galaxies with the same total magnitude, but with different scale lengths. Two isophotal thresholds are shown: the fainter one corresponding to the Ferguson (1989) Fornax cluster survey, and the brighter one corresponding to the APM surveys of Davies et al. (1987) for the Fornax cluster and Loveday et al. (1992) for the field. For the right panel of the figure, imagine that the intrinsic field galaxy LF is described by a Schechter function with $M_B^* = -21$ and $\alpha = -1.5$. Imagine also that the faint end is dominated by galaxies that have exponential profiles (§2.2.1) and obey the following relation between central surface-brightness and luminosity:

$$\mu_0(L) = -2.5\log(L/L^*) + 21.6 \pm 0.5. \qquad (7)$$

(This is basically a constant size relation with some scatter, chosen to be illustrative rather than to reflect reality.) The right panel shows the intrinsic LF and the LF recovered by a field-galaxy survey similar to that of Loveday et al. (1992). The recovered shape is significantly flatter than the intrinsic slope both because many galaxies fall below the isophotal threshold, and because the constant correction from isophotal to total magnitudes systematically underestimates the flux from the galaxies with the lowest surface brightnesses (McGaugh 1994; Ferguson & McGaugh 1994).

While such surface-brightness dependent selection effects are important, we will argue below that they are probably not sufficient to account for the observed differences between the cluster and field LF's.

For clusters, an additional, equally important assumption is that cluster members can be separated from interlopers, either statistically or through redshift or morphological criteria. As galaxy counts at faint magnitudes rise steeply ($logN(m) \propto 0.6m$), improper background correction can seriously affect the estimate of $\alpha$ in cluster samples. For *nearby* clusters, where the increased surface density of galaxies is not very high, statistical comparison of counts in cluster fields to "control" fields is not very useful. If the angular correlation function is a power law $w(\theta) = A\theta^{1-\gamma}$, then the variance in galaxy counts in a square field of $\theta$ degrees on a side is

$$\Delta N^2 = N + 2.24 N^2 A \theta^{1-\gamma} \qquad (8)$$

where N is the number of galaxies in the field, and $\gamma \approx 1.8$ (Peebles 1975). Consider the case at $B = 20$. Counts are about $200\,\mathrm{mag}^{-1}\mathrm{deg}^{-2}$, and $A \approx 2 \times 10^{-2}$ (Roche et al. 1993). Over a $6° \times 6°$ field, the variance is $\Delta N^2 = 5.6 \times 10^5$, meaning that random $6° \times 6°$ patches of the sky will show an RMS fluctuation of $\Delta N = 750$ galaxies at $B = 20$. For comparison, the morphologically-selected Virgo cluster catalog (Binggeli et al. 1985) has only 112 cluster members with $19 < B < 20$ in a field of roughly the same area. Because redshifts are very difficult to obtain for such low-surface-brightness dwarfs, morphological selection is currently the *only* way to obtain the galaxy luminosity function to faint absolute magnitudes.

The corollary is that if morphological selection does *not* identify the full population of dwarfs in places like the Virgo cluster (e.g. if there are many compact dwarfs that look just like background galaxies or many LSB dwarfs below the thresholds of photographic surveys), then the luminosity function could be steeper than currently thought.

### 5.2. Clusters vs. Field

Comparisons of the luminosity function in high and low-density regions are of great interest, both as a test of biased galaxy formation (see §7.2), and as a measure of environmental influences on galaxy evolution. One of the great difficulties here is in establishing the connection between observations and theory. On the theoretical side, there are many different variants of biasing (e.g. Rees 1985). While statistical bias in the distribution of fluctuation peaks (Kaiser 1984; Bardeen et al. 1986) may plausibly bias the spatial distribution of dark-matter halos, the connection between that distribution and the LF is not well understood (in large part because it involves star formation). On the observational side, the techniques used to probe the LF in clusters and outside of clusters are quite different, leading to different potential biases in the results (see above). Finally, there is the difficulty of relating the variations with position or local density *within* a virialized or partly virialized cluster (e.g. the morphology–density relation) to the expectations from linear theories of structure formation.

With these difficulties in mind, we shall briefly summarize the observations from the richest (but not necessarily the densest) environments, to the poorest, concentrating on the dE contribution.

The richest region that has been surveyed to faint limiting magnitudes is the Coma cluster. Thompson & Gregory (1993) catalogued dwarf galaxies in the Coma cluster, identifying faint cluster members on the basis of photographic $b - r$ colors. They measured a faint-end slope $\alpha = -1.4$, and a dwarf/giant ratio slightly lower than that of Virgo, when converted to the same luminosity limit. There is a pronounced lack of low surface-brightness dE galaxies (classified dSph by Thompson & Gregory) within the central 0.3 degrees. They interpret this as the result of tidal disruption in the core of the cluster of galaxies with velocity dispersions less than $35 \text{ km s}^{-1}$. While not explicitly shown, the lack of LSB dwarfs in the core of the cluster presumably leads to a significantly flatter luminosity function. Outside the core of the cluster, the early-type dwarfs follow the spatial distribution of E and S0 galaxies.

The Virgo cluster was extensively studied by Sandage et al. (1985b). They found an overall faint-end slope of $\alpha = -1.30$ and a dE slope $\alpha = -1.35$ to an absolute magnitude of -11.7. Comparison to the Coma cluster survey is a bit uncertain as the morphological classifications are different and the galaxies are obviously not as well resolved at the Coma distance. In contrast to Coma, there is no strong deficit in the number of faint dE's near the center of Virgo, but there is perhaps a slight decline (see Ferguson & Sandage 1991). Bothun et al. (1991) re-examined the Virgo cluster LF, adding in a sample of 24 extremely low surface-brightness galaxies. They examine only the subset of galaxies for which there is quantitative surface photometry, and apply a rather indirect method of calculating the luminosity function, assuming all galaxies have exponential profiles and deriving the distribution of scale-lengths and central surface-brightnesses in a region of parameter space essentially free from selection effects. Modeling these distributions as power laws, they arrive at an LF slope $\alpha = -1.6$. However, the difference between this result and that of Sandage et al. (1985b) is largely a result of the analysis technique, rather than the inclusion of the additional LSB galaxies.

Ferguson & Sandage (1991) compiled luminosity functions for seven nearby groups (including the previously-published Fornax and Virgo clusters), and found faint end slopes $-1.6 < \alpha < -1.3$, and turnover magnitudes $-23.2 < M_B^* < -21.2$. A Schechter function was typically not a very good fit to the data, but the systematic departures from the data were different for different clusters and difficult to separate from uncertainties in the magnitude estimates and completeness. The most striking result from the survey is that the dwarf/giant ratio *for early-type (E, S0, dE, and dS0) galaxies* varies by a factor of $\sim 5$ from the richest to the poorest groups (see §6.2). This result has been extended to still poorer groups by Vader & Sandage (1991).

Surveys of the field typically find flatter luminosity functions, with $\alpha \approx -1$ (Efstathiou et al. 1988; Loveday et al. 1992; Schmidt & Boller 1994), and similar $M_B^*$. Ferguson (1992a) and Lacey et al. (1993) considered the effect of isophotal selection on the field-galaxy luminosity function, and concluded that surface-brightness selection *alone* could not account for the difference in the faint end slope between the Virgo cluster (Sandage et al. 1985b) and the field (Loveday et al. 1992). The Loveday et al. survey was deep enough that relatively high surface-brightness dE galaxies with $-18 < M_B < -16$ would have been detected. However, if the dwarfs in the general field follow the Virgo cluster dE luminosity function, it is quite possible that the total luminosity function turns up below the limits of the Loveday et al. survey. The loose groups observed by Ferguson & Sandage (1991) show such an upturn.

Although they did not explicitly compute a luminosity function, the sample of Eder et al. (1989) provides some indication that the LF in the field, even when probed to fainter surface-brightness limits, is still dominated by HI rich galaxies, with a luminosity function similar to that of Virgo. 122 galaxies from a well-defined diameter-limited sample were observed at 21-cm and only 28 (22%) were not detected. For those detected, the HI properties are similar to the Virgo cluster irregular galaxy sample. However, in the Virgo cluster dE's outnumber Sd+Im galaxies by at least factor of 3 (the exact value depends on luminosity). The flat LF of the field therefore reflects the flat LF of the Sd+Im types. Jerjen et al. (1992) arrived at a similar conclusion. For a sample of groups within 10 Mpc, they found that the faint population is dominated by irregulars (only 11% being dE+dS0) types, and the resulting LF to $M_B = -15$ was flat with $\alpha = -0.98$

The Local Group provides information on the very faint tail of the field-galaxy luminosity function. Van den Bergh (1992) finds $\alpha = -1.1$, although the number of galaxies is probably too small to argue that this is significantly different from the steeper slope found in clusters. Ferguson (1992a) finds a much steeper slope $\alpha = -1.9$ for the M81 group, although once again the statistics are very poor.

Compact groups in principle provide an interesting comparison to the above studies. Such groups can have densities as high as the centers of rich clusters, but have velocity dispersions and total numbers of galaxies consistent with "loose groups." De Oliveira & Hickson (1991) found a luminosity function significantly different from that of field, loose-group, and cluster galaxies, with $\alpha \approx -0.2$. More recently, Ribeiro et al. (1994) found $\alpha = -0.82 \pm 0.09$ from a deeper survey of a large number of groups. The Ribeiro et al. luminosity function appears to be consistent with that seen in the field and in loose groups, but significantly flatter than that of the Virgo cluster. The result is in accord with the general variation of dwarf/giant ratio with richness (as opposed to density), but is rendered somewhat uncertain by the rather bright isophotal threshold of the Ribeiro et al. survey.

While there are still large uncertainties in the luminosity function and its environmental dependence at absolute magnitudes fainter than $M_B = -16$, we believe that surveys to date have taught us the following:

1. Selection effects are important, but not dominant. Dwarf galaxies are probably not a significant contributor to the mass-density of the universe.

2. The overall luminosity function changes with environment in a way that is consistent with a simple change in the relative proportions of dE and irregular galaxies.

3. The dE luminosity function is largely independent of environment, with the exception of a possible depletion at the faint end in the centers of rich clusters (Thompson & Gregory 1993). The slope in the range $-16 < M_B < -13$ is $\alpha \approx -1.3$. The observations for clusters and groups, even when selection effects are taken into account, do not support slopes as flat as $\alpha = -1$ or as steep as $\alpha = -1.8$.

4. LF variations are not consistent with simple biasing schemes. The relative abundance of low-luminosity galaxies is, if anything, lower in the field than in clusters.

## 6. Spatial Distribution and Clustering Properties

The well-known morphology-density relation for the classical Hubble types (Dressler 1980) holds also for dwarf galaxies (Binggeli et al. 1987; Ferguson & Sandage 1988; Binggeli et al. 1990). It may, in fact, even be stronger for dwarf galaxies: Gas-rich dwarfs appear to be the most weakly clustered, i.e., the most uniformly distributed species of galaxies (Salzer 1989), while early-type dwarfs have been said to be the most strongly clustered of all galaxies (Vader & Sandage 1991). The morphological segregation is plausibly stronger for low-mass galaxies, as these systems will also be more vulnerable to environmental influences than high-mass galaxies. Bright, nucleated dwarf ellipticals are infallible tracers of high-density places; they occur only in clusters of galaxies or, if outside, as close companions to massive galaxies. This is generally true for early-type dwarfs (Einasto et al. 1974; Binggeli 1989; Binggeli et al. 1990). However, it is important to note that there seem to exist some fairly *isolated*, faint, non-nucleated dE's. The prime example is the recently discovered Tucana system, which does not show any young, or even intermediate-age population (da Costa 1994). Another case of a possible isolated faint dE has been reported by Binggeli et al. (1990). This would suggest that faint dwarfs can exhaust their gas also without external pressure (§7.6). It is difficult to give an upper limit to the space density of isolated dE's, first, because of their low visibility, and second, because, once detected, their distances are in general not accessible (at least at present). However, recent deep searches for low-surface brightness objects suggest that there cannot be too many of them (Eder et al. 1989; Phillipps et al. 1994). The best guide we have is the Local Group, where the strong clustering of dE's around M31 and the Milky Way has recently been reinforced by the null result of an extended search for new local dwarfs (Irwin 1994; but see Ibata et al. 1994).

### 6.1. Distribution within Clusters

Dwarf ellipticals are the most numerous galaxy type in clusters of galaxies, if cluster members are counted to sufficiently faint magnitudes. Detailed studies of the projected distribution of dE's so far exist only for the nearby Virgo and Fornax clusters (Binggeli et al. 1987; Ferguson & Sandage 1989). Binggeli et al. (1987) have qualitatively shown that nucleated dE's follow the strongly clustered distribution of E's, with non-nucleated dE's being slightly more spread out. This difference between dE,Ns and normal dE's was noted before by van den Bergh (1986), but a later study (Ichikawa et al. 1988) denied the statistical significance of the effect. However, by differentiating their dwarf sample with respect to luminosity, Ferguson & Sandage (1989) did find a significant difference in the distribution of nucleated dE's and *bright* ($M_B < -14.2$) non-nucleated dE's, reproduced here in Fig. 8. The latter follow the shallow cluster profile of spirals and irregulars, hinting at a possible evolutionary link between bright dE's (no N) and late-type galaxies (§7.6). It has also been suggested that the nuclear strength of dE,Ns might correlate with environmental density (Binggeli et al. 1987, p.259, footnote), but this has never been assessed quantitatively.

Otherwise, the dE population of the Virgo cluster seems well mixed. The only significant correlation between photometric and environmental parameters found by Ichikawa et al. (1988) is that dE's in the central ($r \leq 5°$) cluster region are slightly larger and brighter than those outside ($r > 5°$). This mixed state does not, however, mean that all dwarf ellipticals are freely floating in the cluster potential. Ferguson (1992b) has statistically shown that roughly 7% of all Virgo cluster galaxies are bound companions, i.e. are gravitationally bound to a more massive neighbor. By virtue of their overall predominance in the cluster, the majority of companions (66%, or about 80) are dwarf ellipticals. The primaries tend to be massive, early-type galaxies. Similar conclusions have been reached by Giovanardi & Binggeli (1991) and Binggeli (1993). As expected from the tidal influence by the primaries, the mean surface brightnesses of the dE companions are marginally ($\approx 0.3$ mag) higher than those of free-floating dE's (Ferguson 1992b). Binggeli (1993) identified some of the bright probable companions of the first and second ranked Virgo cluster members, M49 and M87, as inferred from the supposed (dark halo) masses of the primaries.

*6.2. Dwarf Companions in the Field*

Holmberg's (1969) classical, statistical study of satellites around bright spiral galaxies has only recently been superseded by Vader & Sandage (1991), who identify dwarf companions *individually* by morphology, and by Zaritsky et al. (1993), who identify them by kinematics. A spectroscopic follow-up of Vader & Sandage (1991) has been started by Vader & Chaboyer (1992). Vader & Sandage (1991) present first results for E and S0 primaries, most satellites of which turn out to be early types as well, i.e. dwarf ellipticals. The projected distribution of the companions is shown to follow a surface density law close to $\Sigma(r) \approx r^{-1}$, in accord with Zaritsky et al.'s (1993) finding for spiral primaries. Interestingly, this is also the profile of the dark halos of spiral galaxies as derived from rotation curve studies. The correlation function of dE's on the smallest scale (around massive giants) is steeper than for any other type of galaxy.

Vader & Sandage (1991) confirm the trend of increasing early-type dwarf frequency per giant galaxy with increasing richness [3] of the aggregate (absolute number of giants) found by Ferguson & Sandage (1991). *Note that this environmental variation is distinctly different from the morphology-density relation: there is no significant variation of the dwarf/giant ratio within a given cluster, while the spiral/elliptical ratio changes dramatically with radius.* In addition, the percentage of dwarfs that are bound to giants *decreases* with increasing richness: virtually all dE's in the field are bound to giants, while in the Virgo cluster only a few percent of them are (Ferguson 1992b). Tidal stripping is a possible explanation of the low frequency of bound satellites in a cluster environment. In the field, galaxies that start off within about 50 kpc of the primary will be accreted in less than a Hubble time, while in a cluster such galaxies could be "liberated" by the tidal field of the cluster, leading to a higher $d/g$ ratio in clusters. However, it is not yet clear whether such a process could produce the observed correlation of $d/g$ with richness. Other possibilities are discussed in §7.2.

In examining the spatial distribution of nearby swarms of dwarf companions around larger galaxies Einasto et al. (1974; 1975) noticed an intesting segregation of dwarf ellipticals from dwarf irregulars. At fixed luminosity, dwarf ellipticals are found at smaller separations from the primary galaxy than dwarf irregulars. The radius that divides dE's from Im's increases with companion-galaxy luminosity: luminous Im's are found at small separations, while low-luminosity Im's are not. Einasto et al. (1974) used this result to argue for the existence of massive gaseous coronae around galaxies such as the Milky Way that would act to strip irregulars of their gas (§7.6). However, while the general tendency of dE's to cluster around giants, in contrast to irregulars, is quite evident (e.g., Binggeli et al. 1990), the specific segregation line of Einasto et al. (1974) has not been confirmed by Zaritsky (1992) [who lacks the high morphological resolution of the Vader & Sandage (1991) study, however], or by Ferguson (1992b) for satellites within the Virgo cluster.

---

[3] Note that the quantity "richness" used here is not the same (nor as well defined) as that used by Abell (1958), as his definition is not useful for very poor clusters or loose groups. The entire discussion of environmental variations in the morphological mix would benefit from a more precisely defined measure of richness, or a more physically motivated environmental parameter.

The only satellite system where we have the full 3D information on the distribution is of course our own, i.e. the two Magellanic Clouds plus the eight local dE's (or "local dwarf spheroidals"), which all lie within 250 kpc of the Milky Way Galaxy. The possible correlations of $M/L$ and star-formation history with galactocentric distance have been mentioned previously. Another interesting phenomenon is that the distribution of the local dE's around our Galaxy is not isotropic. This is evident from their sky-projected distribution. Lynden-Bell (1976) and, independently, Kunkel & Demers (1977) noticed that several globular clusters and dwarf spheroidals lie within only a few degrees of the great polar circle defined by the Magellanic Stream and some of the High Velocity Clouds. The most closely associated dwarf systems are Sculptor, Draco, Ursa Minor, and Sextans. It is significant that Sextans was only recently discovered (Irwin et al. 1990b) and that no other dwarf has thereafter been found in an extensive survey of the southern sky (Irwin 1994), although one has since turned up serendipitously (Ibata et al. 1994). The causal connection between these dwarfs and the Magellanic system remains unclear. It has been suggested that the dwarfs, like the Magellanic Stream, are tidal debris from a recent close encounter of the Clouds (Kunkel 1979; Murai & Fujimoto 1980; Lin & Lynden-Bell 1982; Gerola et al. 1983). However, with respect to their structure and content, the local dE's are indistinguishable from the dE satellites of M31 (Armandroff et al. 1993; Armandroff 1994) where no phenomenon like the Magellanic Stream is known (although the M31 dwarfs also lie roughly in a plane).

## 7. Evolutionary Scenarios

Table 2 summarizes the phenomena to be explained by any theory of dwarf galaxy evolution. Next to each phenomenon or correlation, we put a score (1-5), indicating our judgement of how secure the observations are. A five indicates that the phenomenon is highly significant statistically and is not due to selection effects; 3 indicates that it may be the result of observational selection, or is only marginally outside the observational uncertainties; smaller numbers indicate that the reported phenomenon is probably not real.

While no theory yet claims to account for all the observed phenomena, the most appealing are those that start from gaseous conditions in the early universe and follow the collapse and cooling of structures within the gravitational hierarchy. Such models (White & Rees 1978; White & Frenk 1991; Cole 1991; Blanchard et al. 1992; Cole et al. 1994; Kauffmann et al. 1994; Lacey & Silk 1991; Lacey et al. 1993) predict the statistical properties of dE galaxies, and may therefore be tested against the global phenomena shown in Table 2. Other models deal with the physical mechanisms of how dE's might have lost their gas (Faber & Lin 1983; Kormendy 1986) how star formation might be regulated (Gerola et al. 1980; Lin & Murray 1992; Silk et al. 1987; Efstathiou 1992), and how at least some dwarfs may have emerged as distinct entities (Gerola et al. 1983; Mirabel et al. 1992). However, such ideas must still be incorporated into a larger theory in order to explain the distribution functions of luminosity, surface brightness, color, nucleation, etc.

Rather than review the models individually, we outline below the physical processes that appear to be important. These are the gravitational collapse of the protogalaxy (§7.1), cooling of the entrained gas (§7.3), various external agents that could suppress or slow down the cooling process (§7.4), various internal agents such as winds or photoionization that could regulate star formation (§7.5), and finally, environmental influences acting at relatively recent epochs that could convert low-luminosity star forming galaxies into dE's (§7.6).

Table 2: dE Phenomena

| Phenomenon | Score | Environment |
|---|---|---|
| Surface-brightness–luminosity relation ($\langle\mu\rangle_e \approx 0.75 M_B + 35.3$) | 5 | Clusters |
| Surface brightness profile changes with $L$ | 5 | Clusters |
| Metallicity–luminosity relation ($<[Fe/H]> \approx -0.2 M_V - 3.9$) | 5<br>3 | Local Group<br>Clusters |
| Widespread existence of intermediate age stars | 5<br>3 | Local Group<br>Clusters |
| High $M/L$ in some dE's,<br>Moderate $M/L$ in others | 4<br>5 | Local Group<br>Local Group |
| $M/L$ vs. $L$ relation ($\log(M/L) \approx 0.2 M_V + 3.6$) | 4 | Local Group |
| Apparent $M/L$ vs. environment relations | 3 | Local Group |
| Velocity anisotropy | 4 | Local Group, Clusters |
| Luminosity Function slope $\alpha \approx -1.3$ | 4 | Clusters, Groups |
| Richness dependence of dwarf/giant ratio ($\log d/g \approx 0.65 \log g - 1.05$ for early-type galaxies with $M_B < -13$) | 4 | Clusters, Groups, Field |
| Correlation of nucleation with luminosity ($N(dE,N)/N(dE) \approx -0.15 M_B - 1.7$ for dE's in the range $-12 > M_B > -18$) | 5 | Clusters |
| Environmental variation of nucleated dE fraction | 5 | Clusters |
| Correlation of nucleation with color | 2 | Clusters |
| Anisotropic spatial distribution around giants | 4<br>3 | Local Group<br>Field Companions |

### 7.1. Gravitational Collapse

The leading paradigm for galaxy formation involves gravitational collapse of primordial density fluctuations, followed by cooling of the gas and star formation. The sucess of the COBE satellite in detecting fluctuations in the microwave background both supports gravitational instability as the mechanism for forming structure, at least on large scales, and rules out competing models such as explosions (Wright et al. 1992; Efstathiou et al. 1992). Blanchard et al. (1992) provide a detailed discussion of the theoretical construct, which leads to a mass function

$$N(M)dM \propto M^{-(2-a)}dM \qquad (9)$$

for scale invariant fluctuations of the form

$$\frac{\delta \rho}{\rho} \propto M^{-a} \propto M^{-(n+3)/6}, \qquad (10)$$

where $n$ is index of the initial power spectrum $|\delta^k|^2 \propto k^n$ (Peebles 1980).

The *mass* function of dwarf galaxies is in principle a sensitive measure of the shape of the primoridial fluctuation spectrum, as it is not very sensitive to the nature of the density fluctuations (Gaussian or non-Gaussian) or the details of nonlinear collapse (Blanchard et al. 1992). However, the analytical theory is difficult to test at the resolution of current numerical codes and the mass function is difficult to derive from observations. The CDM power spectrum has $-3 < n < -2$ on the scale of dwarf galaxies, leading to a mass function $N(M) \propto M^{-2}$. In comparison, the observed *luminosity* function has a faint end slope $\alpha \approx -1$ to the limits of field samples and $\alpha \approx -1.3$ in clusters, and perhaps to fainter limits in the field (see §5).

### 7.2. Biasing

If the formation of galaxies depends on environment, then both the fraction of mass in galaxies and the mass function of galaxies may in the end vary over large scales. The concept of biased galaxy formation emerged in the 1980's (Kaiser 1984; Rees 1985; Davis et al. 1985; Bardeen et al. 1986) as a way to reconcile the theoretically attractive "flat universe" with $\Omega = 1$ to the lower $\Omega$ inferred from the virial analysis of rich clusters and from studies of large-scale structure.

There are many plausible ways to introduce bias into the distribution of galaxies (Rees 1985; Dekel 1986). Spatial variations in the efficiency of galaxy formation could arise from the nature of the dark matter (e.g. if a significant fraction of $\Omega$ is in hot neutrinos), from the existence of a universal threshold in the density required for galaxy formation (see below), or from the coupling of galaxy formation to the feedback of energy from the first generations of objects (e.g. via photoionization of the IGM from early starbursts or AGN, or via bulk flows introduced by early supernovae). Because of the uncertainties in the physics, it has been convenient to parametrize biasing by a single parameter $b$ that is the ratio of the rms density fluctuations in luminous matter to the underlying rms mass fluctuations. Because ellipticals and spirals cluster differently, the value of $b$ obviously depends on galaxy type. The COBE results suggest that $b$ also depends on the scale over which the rms fluctuations are measured (Efstathiou et al. 1992; Padmanabhan & Narasimha 1992), at least for CDM-like power spectra.

Some type of biasing mechanism is clearly necessary to explain the clustering properties of dE galaxies (§6). Perhaps the most thoroughly developed is the concept of *statistical* biasing, based on the idea that galaxies form above some global threshold $\nu$ in the Gaussian random field of primordial fluctuations ($\nu = \delta/\sigma$ where $\sigma$ is the rms density variation on a given scale). A standard hypothesis (Dekel & Silk 1986; White et al. 1987; Schaeffer & Silk 1988) has been that giant galaxies form only at $\nu \approx 2-3$, while dwarfs can form at any peak (and so the vast majority of them should come from typical peaks of $\nu = 1$). Such a model predicts that giant galaxies should be more correlated than dwarfs. The enhancement in the giant-galaxy two-point correlation function $\xi_g$ relative to the mass two-point correlation function $\xi_m$ is approximated by (Kaiser 1984; Politzer & Wise 1984)

$$1 + \xi_g(r) = \exp[(\nu^2/\sigma^2)\xi_m(r)], \qquad (11)$$

which corresponds to roughly an order of magnitude at $\nu/\sigma = 3$. Although an explicit estimate of $\xi_g$ for dE galaxies has not been made, the available data clearly rule out any appreciable decrease in $\xi_g$ from giant E's to dE's (see §6). (The data are still ambiguous for the comparison of field Irr's to field spirals – Eder et al. 1989; Salzer et al. 1990; Alimi 1994.)

West (1993) has argued that the same sort of statistical bias may account for different specific globular cluster frequencies of spiral and elliptical galaxies, and the higher globular cluster frequencies of cD galaxies. If one assumes that globular clusters can only form at density peaks higher than some global critical value $\nu$, the enhancement in the globular cluster frequency in galaxy and cluster halos can be calculated from the formalism described above. For reasonable choices of the power spectrum and the density thresholds for spirals, ellipticals, and rich galaxy clusters, the observed variation of the globular cluster specific frequency can be reproduced. West (1993) speculates that the same reasoning might apply to dwarf galaxies, which for the same threshold $\nu$ ought then to be *more* strongly clustered than globular clusters.

As a mechanism for forming globular clusters, West's proposal faces serious difficulties. For example, there is no evidence that globular clusters have dark-matter halos, as they must in this model. Furthermore, the CDM power spectrum, which produces the best match to the trends in globular-cluster specific frequencies, would predict a globular-cluster mass function steeply rising to low masses, and extending to higher masses, contrary to observation. Finally, it appears that at least some globular clusters, for example the relatively young ones observed in the LMC and NGC1275 (Holtzman et al. 1992), form at late epochs, probably not through collapse onto primordial density fluctuations.

On the other hand, West's proposal is perhaps more attractive for explaining the clustering of dE galaxies, which show strong evidence for dark matter and have a steeply rising mass function (although not as steep as CDM models predict). The idea that statistical bias could *enhance* the clustering of dE galaxies runs contrary to all previous statements about the influence of statisical bias on the clustering properties of dwarf galaxies, but nevertheless has the positive feature that it could potentially account for the variation of the dwarf/giant ratio with cluster richness (§6), and it may provide a natural explanation for the high central *dark-matter* densities inferred for some dwarf ellipticals (§3). However, interpretation of the clustering of dE galaxies in the context of statistical biasing requires that their formation threshold $\nu$ was comparable to that of giant elliptical galaxies. The condition that *giant* ellipticals form at high $\sigma$ peaks arises naturally from the standard cooling time argument: the density of their halos must be high enough that the cooling time will be shorter than the free-fall time (§7.3). However, the cooling time decreases rapidly with decreasing halo velocity dispersion ($t_{cool} \propto V^2$), so this mechanism cannot be invoked to argue for a high $\nu$ for dE galaxies. While other possible mechanisms exist (West 1993), missing from the argument is a compelling physical reason why dE's (or globular clusters) should form only above some global density threshold.

### 7.3. Cooling

In hierarchical models, structure grows by gravitational collapse of succesively larger structures. The emergence of galaxies as separate entities within this hierarchy depends on the ability of baryons within a given overdense region to cool and form stars before being subsumed into the next higher level of the hierarchy. Cooling of gas in protogalaxies is thought to be a runaway process. As soon as gas is able to cool in a galaxy halo it contracts, whereupon the cooling becomes more efficient, leading to more cooling and contraction until something (e.g. star formation) is able to put a halt the process. The virial temperatures ($kT_{vir} \approx \mu m_H \sigma^2$, where $\mu m_H$ is the mean molecular weight of the gas) of collapsing protogalaxies are high enough that the associated gas will be collisionally ionized and cool radiatively. In the absence of other energy sources or sources of ionization, the cooling time at radius $r$ can be written as

$$t_{\rm cool} \approx \frac{3\rho(r)}{2\mu m_{\rm H}} \frac{kT}{n_e^2 \Lambda(T)}, \qquad (12)$$

where $\Lambda(T)$ is the cooling rate due to bremsstrahlung, recombination, and collisionally excited line emission, and $\rho(r)$ is the density at $r$ (e.g. White & Frenk 1991; Cole et al. 1994).

For dwarf galaxy halos collapsing at $z \approx 3-10$, the cooling time is short compared to the free-fall time, so cooling should be efficient. Unless this cooling is suppressed, most of the baryonic material in the universe would collapse into dwarf-galaxy size objects. This is the "overcooling problem" noted by Cole (1991) and others. As dwarf galaxies do not appear to be so overwhelming abundant, some mechanism to counteract cooling is needed. Plausible mechanisms involve both internal and external agents.

### 7.4. Suppression of Star Formation: External Agents
#### 7.4.1. Photoionization

Rapid cooling of proto-dwarf galaxies could be prevented if the gas is kept photoionized by the metagalactic radiation field. Babul & Rees (1992) and Efstathiou (1992) argue that the ionizing background at $z > 1$ is high enough to keep the gas in dwarf galaxy halos confined and stable, neither able to escape, nor able to collapse and form stars. The extent to which this effect is important depends on the shape of the ionizing spectrum and its evolution, neither of which are well quantified. The lack of a detectable Gunn-Peterson Lyman-$\alpha$ absorption trough in the spectra high-redshift QSO's (Steidel & Sargent 1987; Webb et al. 1992) suggests the IGM is highly ionized. The ionizing radiation field estimated from the proximity effect ($J_\nu \approx 10^{21}\,\mathrm{erg\,cm^{-2}\,s^{-1}\,Hz^{-1}}$; Lu et al. 1991) appears sufficient to prevent the gas from cooling in halos of velocity dispersion less than $\sim 35\,\mathrm{km\,s^{-1}}$ until $z \approx 1$. The advantage of the model is that it provides a clear connection between dwarf galaxies and QSO Ly$\alpha$ absorbers (the latter being dwarf galaxies in their latency period before cooling), and it provides a source of faint blue galaxies at $0.5 < z < 1$. If AGN are the dominant source of ionizing radiation, a further prediction of the model is that the spatial distribution of dwarf galaxies could be modulated by the distribution of AGN.

### 7.4.2. Reheating

If AGN do not provide sufficient flux to photoionize the IGM (Shapiro & Giroux 1987), an alternative solution to satisfying the Gunn-Peterson test is to suppose that the intergalactic medium was reheated during the epoch of galaxy formation. Mechanisms include heating by supernova winds from protogalaxies (Tegmark et al. 1993), Compton heating from energetic objects at very high redshift (Collin-Souffrin 1991), or a variety of other possibilities (Blanchard et al. 1992). In any case, if the IGM is maintained at a constant temperature $T_{\mathrm{IGM}}$, the only galaxies that can collapse are those with virial temperatures higher than the temperature of the IGM. Blanchard et al. (1992) argue that this consideration leads to a mass-function slope close to the observed one. Outside of the deep potentials of groups and clusters, the reheated IGM cools adiabatically, with temperature $T \propto (1+z)^2$. The minimum velocity dispersion for a galaxy therefore evolves as $\sigma_{\min} \propto 1 + z$. The mass function of halos in the CDM model scales as

$$N(M,z) \approx \frac{1+z}{M^2}. \tag{13}$$

However, all objects collapsing at a given redshift $z$ have the same density $\rho \propto (1+z)^3$. The velocity dispersion scales as $\sigma^2 \propto M/R \propto \rho^{1/3} M^{2/3}$, so

$$M(z) \propto \sigma^3 (1+z)^{-3/2}, \tag{14}$$

which is proportional to $(1+z)^{3/2}$ when we constrain $\sigma_{\min}$ to follow the IGM temperature. Combining these equations leads to a mass function

$$N(M) \propto M^{-4/3}, \tag{15}$$

consistent with the cluster observations summarized in §5, if $M/L$ is roughly independent of $L$. However, this argument only applies outside of groups and clusters, where the observed luminosity function slope is flat to the limits of the observations. The $\alpha = -1.3$ slope for the luminosity function of cluster dE's must be due to some other cause.

### 7.4.3. Merging and shocks

The epoch of dwarf galaxy formation may also be the epoch of rapid merging, at least for a CDM power spectrum in an $\Omega = 1$ cosmology. Shock heating during the mergers can partially counteract the cooling according to eqn. 12. However, Blanchard et al. (1992) conclude that this effect alone cannot suppress cooling enough to avoid overproducing dwarf galaxies.

### 7.4.4. Instabilities

The standard cooling-time calculation assumes that gas in a protogalaxy starts out in a singular isothermal sphere, in thermal equilibrium with the dark matter. Radiative cooling then proceeds smoothly from the inside out. Reality is unlikely to be so straightforward, and it is probable that cooling takes place in a turbulent, inhomogeneous medium. Gas at $10^6$ K will be thermally unstable and will likely develop into a two-phase medium due to rapid cooling in the densest subclumps (Fall & Rees 1985), combined with heating from the first generation of stars. Murray et al. (1993) explore the effects of Kelvin-Helmoltz instabilities on clouds moving through a hot medium. Such an effect could truncate the galaxy mass function in clusters of galaxies at velocity dispersions $\sigma < 10$ km s$^{-1}$, but is unlikely to have a direct effect on the galaxy mass function at higher masses. Nevertheless, instabilities during the cooling phase may play an indirect role in shaping the galaxy luminosity function by influencing the stellar initial mass function, and hence the number of OB stars and supernovae per unit mass formed.

### 7.4.5. Sweeping

Sweeping of gas by an external medium is a widely cited mechanism for cutting off star formation, and transforming dwarf irregular galaxies into dE's (Lin & Faber 1983; Kormendy 1985; Binggeli 1986). Sweeping is unlikely to have been effective at high redshift. While the mean density of the intergalactic medium was presumably higher – approaching densities of the centers of present-day rich clusters ($n_H \approx 10^{-3}$ cm$^{-3}$) at redshifts $z \gtrsim 7$ – random velocities of galaxies through this medium would have been sufficiently low that stripping timescales would be longer than a Hubble time. Sweeping during the epoch of dwarf galaxy formation is thus not a viable solution to the overcooling problem, although it may nevertheless account for the lack of gas in cluster dE's if some other process does not remove the gas prior to cluster collapse. Sweeping processes are discussed in detail in §7.6.

### 7.5. Suppression of Star Formation: Internal agents

Feedback of energy into the interstellar medium from the first generation of stars in a protogalaxy may regulate subsequent star formation. It seems clear that some feedback is needed. The luminosity–metallicity relation suggests that massive galaxies are able to retain their ISM longer than low-mass galaxies, or had different initial mass functions (although the existence of such a relation remains to be demonstrated for cluster dE's). Feedback from star formation is the simplest mechanism for producing such a relation. However, this feedback can take several forms, and it is not yet clear which one dominates.

### 7.5.1. *Supernovae*

By far the most widely studied mechanism for regulating star formation in dwarf galaxies (or galaxies in general, for that matter) is feedback from supernovae. The simplest criterion for supernova regulation is the hypothesis that the ISM is spontaneously ejected from a galaxy when a fraction $f$ of the cumulative energy of all supernovae ever formed exceeds the binding energy of the remaining gas (Larson 1974). The details of the transfer of kinetic energy from the supernova remnants to the gas are hidden in the parameter $f$. Vader (1986) points out that storage of energy in such models is a major problem; the accumulation of large amounts of hot gas prior to ejection inevitably leads to gas densities so high that rapid cooling ensues. Dekel & Silk (1986) solve the problem by requiring that the SNR's must cover a significant fraction of the volume (i.e. they must overlap) before gas can be ejected. Combining this with the assumption that the star formation rate scales with the free-fall time and the assumption that most of the energy is deposited into the ISM during the Sedov (adiabatic expansion) phase, Dekel & Silk (1986) derive a critical halo velocity dispersion $V_c \approx 100\,\mathrm{km\,s^{-1}}$ as a condition for substantial gas removal. While this result has been widely quoted, the agreement of $V_c$ with the rough dividing line between giant and dwarf ellipticals could be a coincidence, given the uncertainties in the input physics.

Perhaps more important than the existence of a critical velocity dispersion $V_c$ is the conclusion, reached by both Vader (1986) and Dekel & Silk (1986), that mass loss in the form of a *chemically homogeneous* wind in a self-gravitating (as opposed to dark-matter dominated) galaxy cannot simultaneously reproduce both the metallicity–luminosity and surface-brightness–luminosity relation. Dekel & Silk (1986) solve the problem by postulating that the mass loss takes place in a halo made up of such a large fraction of dark matter that the gas loss would have *no dynamical effect* on the stellar system that is left behind. This, combined with an instantaneous recycling approximation for the evolution of metals, leads to a one-parameter model which governs the relation between $L, R, [\mathrm{Fe/H}]$, and $\sigma$. Dekel & Silk (1986) choose the free parameter by *requiring* that the model follow the observed radius–luminosity relation $L \propto R^4$ (equivalent to the surface-brightness–luminosity relation). The success of the model is that, having fixed this one parameter, it can match the observed luminosity–metallicity relation (which Dekel & Silk quote as $L \propto Z^{2.7}$) and the zero-point of the surface-brightness–luminosity relation. The largest discrepancy with observations is the large $M/L$ predicted for dE's in the mass range of Fornax or Sculptor. One of the biggest departures of the model from standard lore is the assumption that stars can form in a gas cloud that is not self-gravitating.

Vader (1986) adopts a different approach. Assuming constant $M/L$ and gas removal that is slow compared to the crossing time of the system in a self-gravitating halo, she arrives at a $r_e$–$\sigma$ relation for dE's that is inconsistent with the observations. Dekel & Silk went through the same argument and arrived at the same conclusion. However, rather than assuming that galaxies are not self-gravitating, Vader assumes that the efficiency $\epsilon$ with which SN energy is converted into gas escape energy is a strong function of the velocity dispersion of the galaxy, with dwarf galaxies having *lower* efficiencies than giants or globular clusters. The basis for this conclusion is the assumption of a constant *initial* mass–density relation for globular clusters, dE's and giant E's. Vader (1987; 1987) goes on to consider the different conditions for wind removal in different environments and the conditions under which the metallicity of the wind will be enhanced relative to the overall ISM, driving out more metals per unit mass than might be expected in models with uniform winds. The model is able to explain why dE's apparently have lower metallicities for their velocity dispersions than giant ellipticals (but see Bender et al. 1993), but can only qualitatively reproduce the position of dE's in the $r_e$–$\sigma$ plane.

Yoshii & Arimoto (1987) argue that both the relations of metallicity and surface brightness with luminosity *can* be reproduced with a chemically homogeneous wind model with no dark matter. They point out that models generally predict a mass-weighted metallicity, while the observations measure a luminosity-weighted metallicity. The latter can be lower than the former by up to 1 dex. Yoshii & Arimoto (1987) construct a model based on the assumption that the binding energy of elliptical galaxies scales as $M^{1.45}$ and the star-formation timescale in the mass range of dwarf galaxies scales as $\tau_{SF} \propto M^{-1/3}$ (from consideration of free-fall time and the collision time of fragmentary clumps). The different structural properties of E's, dE's and globular clusters are determined by the balance between the star-formation timescale and the timescale for the formation of a supernova-driven wind. Winds in proto-dE galaxies set in at $\sim 10^7$ yr, but are slow enough that galaxies can respond to the mass loss, and hence lose more than half their total mass without becoming unbound (however, see Angeletti & Giannone (1990) for a comment on the specific numerical estimate of the time of occurrence of a global wind). Expansion as a result of mass loss accounts for the low dE surface brightnesses. Mass-to-light ratios in this model are expected to be $M/L_B \approx 5-6$ for dE galaxies. The high $M/L$ values for the Draco and UMi dwarfs are thus problematical.

In an ambitious attempt to explain the excess faint blue galaxies, Lacey et al. (1993) propose a model of tidally triggered galaxy formation. Their model for dwarf galaxy formation is similar to that of Dekel & Silk, in that they assume star formation ceases when the gas is ejected by supernova-driven winds. However, they assume that stars form in the baryon-dominated cores of the dark-matter halos, and therefore that the baryonic cores expand homologously when the mass is ejected. They are able to reproduce the surface-brightness–luminosity relation extremely well, but do not consider chemical evolution, and therefore cannot reproduce the metallicity–luminosity relation. Their prediction for B-V vs. mass is completely contrary to observations (more massive galaxies are bluer – although the colors in this model are due to star-formation histories rather than metallicity).

Other efforts to evolve galaxies in hierarchical models adopt a quite different prescription for how supernovae regulate star formation (White & Frenk 1991; Kauffmann et al. 1994; Cole et al. 1994). In these models, heating of the gas by supernovae and subsequent cooling to form more stars reaches equilibrium, thus governing the rate of star formation in the sense that low-mass galaxies have lower star formation rates. The assumption that SN heat the gas rather than eject it is based on the argument that the time between star formation and energy injection is short compared to the sound crossing time or the cooling time. Hence once star formation starts it quickly becomes self-regulating. The star formation timescales $\tau_{SF}$ for dwarf galaxies in these models are long. For example, a typical galaxy with velocity dispersion $\sigma = 50$ km s$^{-1}$ in the Cole et al. (1994) model has $\tau_{SF} = 29$ Gyr. Clearly the dwarfs in these models are dwarf irregulars, rather than dE's. The models do have the advantage that they can schematically reproduce the luminosity–metallicity relation. No attempt has yet been made to match the surface-brightness–luminosity relation. Perhaps the most problematical observation for such models is the apparent lack of rotational support in dE's (§3.2). It is not clear how a self-regulated, mostly gaseous galaxy could remain anisotropic for several billion years.

Athanassoula (1994) describes the results of hydrodynamical simulations of dE formation that include feedback from supernovae. The simulations reproduce the $\mu - L$ relation and the range of dE surface-brightness profiles reasonably well. The models match the observations much better when dE's are assumed not to have dark-matter halos. For the self-gravitating models, the model locus in the $\mu - L$ plane is actually somewhat narrower than the observational data. However it is important to note that each model point at fixed magnitude represents a different set of *physical* assumptions (i.e. initial density profile, coefficients governing cooling, star formation, and supernova feedback). What is remarkable is that varying these parameters does not fill in the $\mu - L$ plane. Evidently the details of cooling, star formation, and feedback are unimportant for producing the surface-brightness–luminosity relation. No attempt has yet been made to match the metallicity–luminosity relation or the variation of profile shape with luminosity for these models.

De Young & Heckman (1994) consider the effect of geometry on the ability of supernovae to clear a galaxy of its ISM. If the galaxy is non-spherical, the wind may clear holes along the minor axis and then be very inefficient at clearing out the rest of the galaxy. De Young & Heckman (1994) find that galaxies of $10^7 M_\odot$ can easily remove their entire ISM, while galaxies of $10^{11} M_\odot$ are very resistant to disruption. Galaxies with masses $\sim 10^9 M_\odot$ show the widest range of behavior. Such an enhanced sensitivity to geometry might help account for the scatter (if real) in the metallicity–luminosity relation of dE galaxies in this mass range.

*7.5.2. OB star winds*

If the initial mass function (IMF) extends to high masses ($\gtrsim 60 M_\odot$), the energy and momentum deposited into the ISM from O and B stars may be comparable to that deposited (at later times) by supernovae. Leitherer et al. (1992) have constructed detailed models of the effect of OB star winds for a variety of assumptions for metallicity and IMF. In a typical case, OB stars dominate both the energy and momentum flux until ages $t \gtrsim 5 \times 10^6$ yr. Thereafter, SN dominate. The total contribution from OB star winds depends critically on the IMF and the metallicity. For constant IMF, lowering the upper mass cutoff of the IMF from $120 M_\odot$ to $30 M_\odot$ decreases the energy output of winds by nearly 2 orders of magnitude. For constant IMF, the energy input increases by 1.5 orders of magnitude from $Z = 0.1 Z_\odot$ to $Z = 2 Z_\odot$. In contrast, the contribution from supernovae is very insensitive to the upper mass cutoff and metallicity. OB star winds, therefore, offer a possible mechanism for introducing variations in the star-formation feedback efficiency as a function of galaxy metallicity, and hence luminosity.

*7.5.3. OB star photoionization*

Most of the energy output by OB stars during their main-sequence lifetime comes in the form of UV photons. This energy ($\sim 10^{53}$ erg over the life of a $30 M_\odot$ star) exceeds that of the final supernova by roughly two orders of magnitude. As with winds, the energy input from this radiation comes well before supernovae become important. Lin & Murray (1992) studied the effect of OB star photoionization on a collapsing protogalaxy and concluded that it could regulate star formation, holding the gas at a temperature $T \approx 10^4 K$, with gas cooling and stars forming at a rate sufficient to keep the gas marginally ionized. Until supernovae become a significant source of energy (at ages $\sim 10^7$ yr), this self-regulation will cause stars to form over timescales longer than a free-fall time, and with different rates as a function of radius in the galaxy. With this self-regulation Lin & Murray (1992) are able to reproduce the $r^{1/4}$ and exponential surface brightness profiles of ellipticals and disk galaxies, respectively, and the Faber-Jackson and Tully-Fisher scaling relations between luminosity and kinematics. The model does not naturally produce exponential surface-brightness profiles for non-rotating galaxies, and therefore may not provide the best explanation for the formation of faint dE galaxies. Nevertheless, photoionization from OB stars may still be important in governing their early star-formation history.

### 7.5.4. *Chemodynamical Models*

Burkert et al. (1994) emphasize that cooling and feedback from star formation are highly sensitive to metallicity. Winds from OB stars, photoionization, dust and the rate of $H_2$ formation, the shape of the IMF, and the minimum mass of type-II supernovae all to some degree depend on metallicity. Hence, processes that were unimportant in the balance between heating and cooling during the early collapse phase of a galaxy may gain in importance as the metallicity of the system increases. Burkert and collaborators attempt to incorporate metallicity dependence into a hydrodynamical model of galaxy formation. For galaxies less than $10^{10} M_\odot$ in these models the effect of feedback is to slow the star-formation timescales such that $\tau_{\rm SF}$ is much longer than a free-fall time. However, in contrast to the models of White & Frenk (1991) and Cole et al. (1994), the variation in feedback efficiency with metallicity allows a galaxy that was not initially able to eject its ISM to do so after a few Gyr. Gas later shed from evolved stars cools and sinks to the center, allowing subsequent generations of stars. In contrast to models involving rapid expulsion of gas (e.g. Dekel & Silk 1986), dE galaxies in these models have small surface densities because they never experience a dissipation and collapse phase. The gas which is lost in the wind has a small fraction of the total mass, but contains most of the metals (as in Vader 1987). Although this has yet to be explicitly demonstrated, such models may be able to match simultaneously both the surface-brightness–luminosity and metallicity–luminosity relations. The initial extended period of star formation that is a robust feature of such models may be tested with deep HST color–magnitude diagrams of the local dE's.

While there is clearly much work to be done to bring the models and observations to the point where they can be quantitatively compared, we conclude this section by suggesting that *the correlations of surface brightness and metallicity with luminosity tend to favor internal over external agents for suppressing star formation, and slow as opposed to rapid mechanisms for gas removal.* If gas removal is indeed slow, then external mechanisms for triggered star formation and/or gas removal, acting well after the time of halo collapse, may greatly influence dwarf galaxy evolution as a function of environment.

## 7.6. Secular Evolution of Irregulars to dE's

Dwarf elliptical galaxies clearly once had gas and formed stars, and therefore must at some time in their past have looked like dwarf irregular galaxies. Could dE's simply be highly evolved irregulars? The secular evolution of irregulars to dE's, with evolutionary time scale being controlled by the environment, could plausibly account for the different clustering properties of dE and irregular galaxies (see §6). There are, in fact, many intermediate-type (dE/Irr) dwarf systems in the Virgo cluster and around massive field giants where we may be witnessing the conversion of an irregular to a dE (Sandage & Binggeli 1984; Vigroux et al. 1986; Sancisi et al. 1987; Sandage & Hoffman 1991; Sandage & Fomalont 1993). In this section we review processes that could bring about such a conversion.

### 7.6.1. Sweeping

Gas will be stripped from a galaxy moving through a hot medium if the ram pressure exceeds the restoring force per unit area due to the galaxy's own gravity (Gunn & Gott 1972). For a spherical galaxy, ram pressure will be effective if

$$\rho_h v^2 > 4\rho_g GM/3R \tag{16}$$

where $\rho_h$ is the density of the hot gas, and $v$, $M$, $R$, and $\rho_g$ are the velocity, mass, radius, and mean gas density of the galaxy (Takeda et al. 1984). Complete stripping of the central regions requires that $\rho v^2$ exceed the maximum binding force, which for a King model is $\approx 3.5\rho_g \sigma^2$, where $\sigma$ is the galaxy velocity dispersion. To get a rough idea of the importance of ram-pressure sweeping, consider a galaxy with $\sigma = 25 \text{ km s}^{-1}$ and a density $n_H = 1 \text{ cm}^{-3}$ moving through an external medium at $700 \text{ km s}^{-1}$. According to the above formula, complete sweeping would require a density in the external medium $n > 4 \times 10^{-3} \text{ cm}^{-3}$, something that is achieved only within the central 100 kpc of the Virgo cluster core. In addition to ram pressure, evaporation, turbulent viscous stripping, and laminar stripping will also act to remove the ISM from an infalling dwarf galaxy (Cowie & Songaila 1977; Nulsen 1982). Nulsen (1982) estimates a typical mass loss rate of

$$\dot{M}_{\text{typ}} = \pi r^2 \rho_h \sigma \tag{17}$$

for a galaxy of radius $r$ in a cluster of velocity dispersion $\sigma$ with a gas density $\rho_h$, for all of these transport processes (different ones acting in different regimes of galaxy velocity relative to the sound-speed in the external medium). In more convenient units, this becomes

$$\dot{M}_{\text{typ}} = 7.4 \times 10^{-2} M_\odot \text{yr}^{-1} \, n \, r^2_{\text{kpc}} \, \sigma_{\text{km s}^{-1}}, \tag{18}$$

where $r_{\text{kpc}}$ is the galaxy radius in kpc, which suggests a stripping timescale of $2 \times 10^9$ yr for a galaxy with a mass in gas of $M = 10^8 M_\odot$ in a cluster with mean gas density $n = 10^{-4} cm^{-3}$ and velocity dispersion $\sigma = 700 \text{ km s}^{-1}$. Such densities are attained within 300 kpc of M87 (Fabricant & Gorenstein 1983). (Timescales are significantly shorter for the Coma cluster.)

The above arguments suggest that in clusters such as Virgo, gas removal will be relatively slow except in the central regions of the cluster. Galaxies on orbits that do not pass through the central $\sim 0.5$ Mpc will probably retain gas for several times $10^9$ years. Galaxies that orbit beyond the central Mpc could retain their gas for a Hubble time. Unless the orbits are predominantly radial, we should expect to see a large variation in the gas content of dwarf galaxies over the central few Mpc of the Virgo cluster. Observationally, dE's are the dominant dwarf-galaxy population over this entire region, and there are no obvious systematic variations in their properties with position in the cluster, other than the frequency of nucleation (§6).

We conclude, purely on the basis of these arguments, that stripping was probably not the dominant gas-removal mechanism in environments such as the Virgo and Fornax clusters. A similar argument suggests that stripping is probably not the major gas removal mechanism for the Milky Way companions. For a galaxy of radius 1 kpc to lose $2 \times 10^7 M_\odot$ in less than 10 Gyr would require $n > 10^{-4}$cm$^{-3}$, averaged over the orbit of the galaxy. ROSAT observations probably exclude $> 10^6$K gas at the required densities for normal spiral galaxies (Pietsch 1992).

However, these simple arguments neglect the fact that the interstellar media of galaxies are multi-phase. Low-density diffuse gas will be much more easily stripped than the high-density gas that is directly associated with star formation. What effect this has on the evolution of a galaxy therefore depends on how gas circulates from dense to diffuse phases — a process that is stopped by sweeping at densities much lower than required for removal of the entire ISM. If sweeping by an external medium *is* the dominant gas-removal mechanism (i.e. if all dE's pass through high density regions at some time during their lives), the natural consequence is that the metallicity–luminosity relation should not exist, or should be modulated by environment. While the stripping timescale depends on galaxy mass, leading to some correlation, it also depends on galaxy orbit and the time when the galaxy encountered a dense external medium, both of which are probably independent of mass, but may introduce correlations of metallicity with position. No such correlations have been identified, but they have not been ruled out either.

Independent of this discussion, there are four classical arguments against a pure stripping scenario for dE's brighter than about $M_B = -14$ (Davies & Phillipps 1988; Binggeli 1994a): (1) bright dE's have nuclei and irregulars do not, (2) bright dE's have surface brightnesses that are too high compared to irregulars (see Fig. 3), (3) bright Im's are flatter than bright dE's, (4) dE's are more metal-rich than Im's (Zinnecker & Cannon 1986; Thuan 1985). Of these, argument (3) now appears considerably weaker (§2.2.3), and argument (2) has not been demonstrated explicitly for complete samples, controlling for the underlying luminosity–metallicity correlation.

Even if interaction with a surrounding medium does not remove all the gas, it is possible that it influences the star formation histories of dE galaxies in other ways. Episodic encounters of dE's with dense gas, either in galaxy halos or in clusters, could provide a mechanism for forming multiple, distinct generations of stars, thus enhancing surface brightness and metallicity, and for forming the nuclei. In this regard, the effect of *pressure confinement* by the external medium may be more important than the gas removal processes discussed above.

*7.6.2. Pressure confinement*

Pressure confinement by the intracluster medium has been considered by Fabian et al. (1980) and by Babul & Rees (1992) in the context of dwarf galaxies. For a gas cloud of mean density $n_g$ and temperature $T_g$ to be in pressure equilibrium with a surrounding medium of density $n_h$ and temperature $T_h$, we require that $n_g T_g \approx n_h T_h$ (ignoring gravity). A galaxy with a $T_g = 100$ K ISM of star-forming density $n_g \approx 1\,\mathrm{cm}^{-3}$ encountering a medium of $T \approx 4 \times 10^7$ K (the approximate temperature of the Virgo cluster gas found by Fabricant & Gorenstein 1983), will be pressure confined at densities $n_h \approx 10^{-6}\mathrm{cm}^{-3}$. A galaxy falling into the cluster for the first time at the present epoch encounters such densities at a relatively low (subsonic) velocity, and may consequently have its ISM compressed long before ram pressure and other gas removal mechanisms become efficient. If the star-formation rate scales with some power of the ISM density, such pressure confinement provides a natural explanation for the existence of blue compact dwarf galaxies (dwarf galaxies with enhanced central star formation) at the peripheries of clusters (Hoffman et al. 1989), and also provides a mechanism for converting Irr to higher-surface-brightness dE galaxies through a period of enhanced star formation prior to gas stripping, similar to the scenario proposed by Davies & Phillipps (1988).

On the other hand, White & Frenk (1991) and Cole et al. (1994) argue that star formation will be self-regulated (see §7.5.1), with dwarf galaxies maintaining a large reservoir of gas at the virial temperatures of their halos by injection of energy from a roughly constant of star formation. Interaction with the ICM will increase the density, and hence shorten the cooling time of the gas. Once the external pressure dominates, the cooling time scales roughly as $t_{\mathrm{cool}} \propto n_h^{-1}$ for a galaxy moving through an isothermal cluster halo. The effect once again will be to increase the star-formation rate during infall into the cluster.

*7.6.3. Tidal Shaking*

Since cooling and star formation (presumably) depend on density, in principle any mechanism that can perturb the density of gas in a quiescent dwarf irregular can increase the star formation rate. If gas is subsequently expelled either by winds or stripping, the galaxy can complete its transition from an irregular to a dE. One possible mechanism that does not require an external gaseous medium is "tidal shaking" (Miller 1988). Tidally induced star formation has been invoked in the context of hierarchical models (Lacey & Silk 1991; Lacey et al. 1993), where the details of the mechanism were not specified. The process has never really been quantified for dwarfs. However, Icke (1985) carried out hydrodynamical simulations for rotating disk galaxies that indicate that even distant encounters can cause shocks in the ISM without significantly perturbing the stars. The development of shocks depends on the mass of the perturber and the mass ratio of the perturber to the perturbed galaxy. Icke's scaling relations imply that for a perturber/perturbed galaxy mass ratio of 1000, shocks will develop if perigalacticon is less than $\sim 30$ perturbed galaxy scale radii. For dE scale radii of $\sim 1$ kpc, this requires rather close (i.e. rare) passages, and is therefore probably not very effective on the outskirts of clusters (where pressure induced star formation might be). However, shocks per se may not be necessary to influence the star-formation rate, which in some models is determined by a delicate balance of cooling and supernova heating. Tidal shaking may be enough to superimpose on this feedback cycle an environment-dependent modulation in the star formation efficiency, which may help explain the environmental variations in the dE fraction, and the episodic bursts of star formation seen in local dE's.

However, we must emphasize that mechanisms such as pressure induced star formation and tidal shaking can only provide a viable way to form the large dE populations of clusters if a suitable reservoir of galaxies exists. The mass function of field dwarf irregulars and HI clouds must therefore have roughly the same slope as the dE mass function. The data are really too sparse to tell whether this is the case. The comparison of the dE and Irr *luminosity* functions certainly does not support such a transition, as the Irr LF appears to be flat (Binggeli et al. 1988). The slope of the HI mass-function from blind surveys is perhaps somewhat steeper (Kerr & Henning 1987; Weinberg et al. 1991; Schade & Ferguson 1994).

An obvious consequence of induced star formation is that dE galaxies may have formed many of their stars *after* the collapse of galaxy clusters. As this occurs rather late in hierarchical models, dE galaxies should contain significant components of stars with ages less than $\sim 5$ Gyr.

### 7.6.4. *Stochastic Star Formation*

A final link between dwarf irregulars and dwarf ellipticals is the possibility that star formation in gas-rich dwarf galaxies is stochastic, with short bursts followed by long periods of dormancy (Gerola et al. 1980; Tyson & Scalo 1988). Such a mechanism is not viable as an explanation for most dE's because they should have large reservoirs of HI. However, only a few dozen dE's have actually been searched for HI, leaving open the possibility that *some* could be dormant dwarf irregulars. Particularly attractive candidates are bright *non-*nucleated dE's, which show the same spatial distribution in clusters as the dwarf irregulars (see §6.1). Only a few have been observed in HI, and none detected.

### 7.7. *dE's as Debris*

Of the other proposed mechanisms for forming dwarf galaxies, the most popular involve debris from giant-galaxy collisions (Gerola et al. 1983; Barnes & Hernquist 1992; Mirabel et al. 1992; Athanassoula 1994). It is no doubt true that galaxy interactions can produce low mass gas clouds, which can subsequently cool and form stars. The question is whether they will look anything like dE galaxies after star formation ceases. Massive dark halos are not expected in this scenario. The strong observational evidence for high $M/L$ in some of the local dE's therefore argues against interactions as the sole formation mechanism. However, direct evidence for dark matter is lacking for most dE's. More indirectly, if Dekel & Silk (1986) are correct in their argument that the metallicity–luminosity–surface-brightness relations *require* dark matter halos, then collisions could not have produced most of the dwarfs. However, it is not obvious that dark halos are required if metal-enriched winds can develop (Vader 1987; Burkert et al. 1994). If so, there is essentially nothing to rule out formation of *most* dE galaxies from tidal debris. Indeed, such a possibility is somewhat attractive in that it may help account for the high dE content of clusters, where presumably many gaseous encounters occured during E galaxy formation (when relative velocities were presumably low). However, while hierarchical models provide detailed (albeit perhaps already falsified) predictions for the dwarf-galaxy luminosity function, tidal debris models have not yet been developed to the point where such predictions can be made.

It is possible that (some) cluster dE galaxies arise as debris of another sort: remnants from the destruction of spiral galaxies in the centers of clusters (Whitmore et al. 1993). The similarity of the bulges of late-type galaxies and dE's suggests that an Sc that has lost its disk might look like a dE. However, assuming the morphological mix in the field represents the initial proportion of giant spirals to giant ellipticals, only a small fraction of the dE's could have been produced this way as the reservoir of spirals is too small.

## 8. Applications to Cosmology

### 8.1. Dwarf Ellipticals as Distance Indicators

Although they are faint, by sheer dint of numbers dE galaxies offer an attractive source for estimating the distances, or at least the relative distances, to nearby clusters. The association of most of the dE galaxies with the relaxed cores of clusters seems unambiguous, thus minimizing the uncertainties in cluster membership that hamper distance estimates based on the Tully-Fisher relation for late-type galaxies. Furthermore, it appears possible to derive purely *photometric* distance indicators for dE galaxies, which may somewhat atone for the difficulty of observing them.

The surface-brightness–luminosity relation has been used by Caldwell & Bothun (1987) and Ferguson & Sandage (1988) to estimate the relative distances to the Virgo and Fornax clusters. The scatter in this relation is much larger than for the $D_n - \sigma$ or Tully-Fisher relations, the deviation from a linear fit to the $\mu_e$ vs. $B_T$ relation being 0.8 mag for the Ferguson & Sandage (1988) Fornax cluster sample. However, the mean of the relation can be determined to comparable accuracy due to the increased numbers of galaxies. Caldwell & Bothun (1987) derived a Fornax/Virgo distance ratio of 0.8 from the $\mu_0$–$B_T$ relation, while Ferguson & Sandage (1988) arrived at a ratio of 1.0, from consideration of the $\mu_e$–$B_T$ relation, and suggested that the previous result may have been biased due to the selection of galaxies in the Virgo cluster photometric sample.

Bothun et al. (1989) estimated relative distances to the Virgo, Fornax, and Centaurus clusters using a subset of dE galaxies with well-defined exponential profiles, and assuming that the scale-length for such galaxies is constant. Later work by Bothun et al. (1991) suggests that the distribution of scale-lengths for dE galaxies in general is a steeply rising power-law ($N(\alpha) \propto \alpha^{-2}$). While this does not explicitly disprove the claim that dE's with pure exponential profiles all have the same scale length, it raises the possibility that distance estimates based on constant $\alpha$ could be subject to strong selection biases, for example based on the criteria used to decide whether galaxies are well fit by exponentials.

The increasing curvature of dE surface-brightness profiles with luminosity (Binggeli & Cameron 1991) provides another possible distance indicator. Young & Currie (1994) parametrize dE profiles with a generalized de Vaucouleurs law $I(r) = I_0 exp[-(r/\alpha)^n]$, and find a reasonable correlation of $n$ with total magnitude for the Caldwell & Bothun (1987) Fornax cluster sample. Based on comparison with four local dE's, they estimate a distance of 13.8 Mpc for the Fornax cluster. The major difficulty in using such a relation to estimate distances is probably the ability to determine $n$, and the fact that $n$ and total magnitude are highly correlated. The authors estimate that they can determine $n$ to better than $\pm 0.06$, but this estimate is based on the typical shot noise in the 1-d profiles and the uncertainty in the sky determination, rather than on a proper reminimization $\chi^2$ over all the other relevant photometric parameters (galaxy center, orientation, ellipticity, central surface brightness, and scale length) for different assumed values of $n$.

All of the above distance indicators are sensitive to selection biases, both in the standard Malmquist sense, and in the sense that most of the existing samples of dE galaxy CCD photometry are biased by surface-brightness in one way or another (see §4.3). Reasonably complete and unbiased samples could be constructed brighter than $M_B \approx -13$ for the Virgo and Fornax clusters, but more distant clusters will be progressively affected by interlopers, such that dE distance indicators may lose their usefulness beyond $v \approx 5000 \, \mathrm{km \, s^{-1}}$.

To obtain *absolute* distances using dE's requires local calibrators. For an assumed distance of 22 Mpc, the extrapolation of Virgo cluster surface-brightness–magnitude relation does not line up perfectly with the Local Group dE's, suggesting a closer distance for Virgo. However, it is difficult to place the resolved Milky Way companions on the same scale due to uncertainties in their photometry. Young & Currie (1994) used NGC 205, NGC 185, NGC 147, and the Fornax dwarf for calibration, and also found a result favoring a short distance scale. While such distances cannot be taken too seriously for the reasons cited above, *in principle* there are actually more local calibrators for such photometric distance indicators than for other techniques, e.g. the Planetary Nebula luminosity function (Jacoby 1989), or surface-brightness fluctuations (Tonry & Schneider 1988). The difficulty is in getting precise photometry, and in providing a convincing physical argument for why the photometric correlations of dE galaxies in clusters should be the same as those in the Local Group.

Finally, dE galaxies may offer useful targets for distance estimates based on surface-brightness fluctuations. The amplitude of the fluctuations is actually greater for dE galaxies than for giant ellipticals, due to the lower surface density of stars and the lower metallicity (Bothun et al. 1991). However, the required exposure times are still much longer for dE's. Counteracting that is the availability of local calibrators, *of the same luminosity and color as the more distant target galaxies* for which fluctuations can be measured directly and computed from color-magnitude diagrams. This may circumvent, or at least provide a check on current absolute distances based on a calibration to M31, M32, and NGC 205 (Tonry 1991).

### 8.2. *Contribution to Faint Galaxy Counts*

The secondary star-formation episode in some of the Local Group dE's could have extended to redshifts as low as $z \approx 0.3$. This suggests a possible link to the faint blue galaxy population that shows up in deep surveys. Several authors have suggested that dwarf galaxies form many (if not all) of their stars at late epochs (Silk et al. 1987; Babul & Rees 1992; Efstathiou 1992). Because of the uncertainties in the physics of star formation, these models do not make very detailed predictions. However, there are a few reasonably model-independent requirements that must be met if dwarf galaxies are to account for the majority of galaxies fainter than $B \approx 21$.

At high redshifts, the $B$ band samples the rest-frame UV and depends more on the star-formation rate than on the number of stars in the galaxy. At a redshift of $z \approx 0.4$, a $B$ magnitude of 24 corresponds to a star-formation rate of 5 $M_\odot$ yr$^{-1}$, for a Salpeter IMF. Even for a large dwarf galaxy, say $10^9 M_\odot$, forming all its stars at a rate of $5 M_\odot yr^{-1}$ would require a relatively short burst of $\sim 2 \times 10^8$ year duration. Less massive galaxies would require even shorter bursts. For comparison, the interval ($0.3 < z < 1$) incorporates a range of 4 Gyr in lookback time. This means that if dwarf galaxies are to supply the excess blue counts, they must form their stars in short bursts over a wide range of redshifts.

The widely varying star-formation histories of the Local Group dE's are perhaps consistent with this requirement, although galaxies even as bright as the Fornax dE are much lower in luminosity than those detected in the deepest redshift surveys. Cowie et al. (1991) estimate luminosities of $0.01L^*$ in the K band for the bulk of the galaxies at $B = 24$. This is closer to SMC luminosities than Local Group dE luminosities. However, the dE sequence in clusters continues up to $\approx 0.05L^*$, so it may be that the lower luminosity dE's are just not showing up to the limits of the current redshift surveys. More problematical is that for the local dE's, the data suggest the star formation in the secondary episodes took place over a few Gyr (Mighell & Butcher 1992; van den Bergh 1994), star-formation rates lower than required by several orders of magnitude.

## 9. Conclusions

Dwarf elliptical galaxies clearly have much to offer for studies of cosmology. The uniformity of their structure and their ubiquity in the local universe make them promising test particles and distance indicators. Correlations among their structural properties and their stellar populations offer clues to galaxy formation that we have yet to fully understand.

For the future, on the observational side we suggest that more detailed quantitative studies of the structure and stellar populations of cluster dE's are vital if we are to provide serious constraints on models for their formation. Samples must be chosen to represent the full range of surface-brightness, morphology, and position in a cluster before we can have a clear idea which parameters are most important. Detailed studies of the stellar populations and luminosity function of the nuclei are also essential if we are to understand how they fit into the process of galaxy formation.

On the theoretical side, the most promising models for explaining the surface-brightness–metallicity–luminosity relations appear to be those with slow, self-regulated star formation, and metal-enhanced winds. Environmental effects such as sweeping or triggered star formation may act in addition, but probably do not dominate the evolution of most dE's. However, some sort of environmental influence seems essential to explain the clustering properties of dE's. The most promising ideas appear to be dE formation from tidal debris as a way of accounting for the high dwarf/giant ratio in clusters, and pressure-induced star formation as a way of accounting for the central concentration of nucleated dE's within clusters. However, much work remains to be done before such ideas can even be tested against the luminosity function, clustering properties, or frequency of nucleation of dE galaxies.

We would like to thank numerous colleagues for providing us with preprints, reprints, references, and especially ideas for this review. We are particularly grateful to Patricia Vader, Brian Chaboyer, Ortwin Gerhard, and Sidney van den Bergh for allowing us to reproduce their figures. This work was supported in part by NASA through grant #HF-1043 awarded by the Space Telescope Science Institute which is operated by the Association of Universities for Research in Astronomy, Inc., for NASA under contract NAS5-26555, and by the Swiss National Science Foundation. This research has made use of the NASA/IPAC Extragalactic Database (NED) which is operated by the Jet Propulsion Laboratory, Caltech, under contract with NASA.

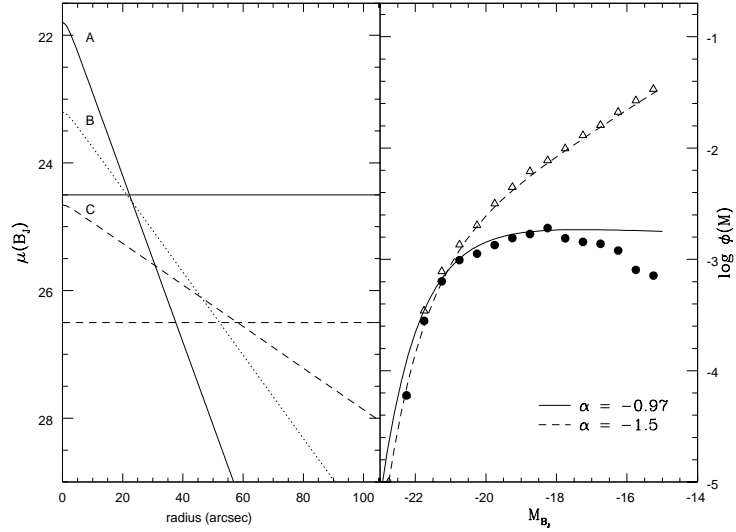

Figure 7:
Illustration of the effect of isophotal selection and isophotal magnitude estimates on the field-galaxy luminosity function. The left panel shows simulated radial surface-brightness profiles for three exponential galaxies. Galaxy A has the canonical Freeman central surface brightness $\mu_0(B_J) = 21.6$. Galaxies B and C have scale lengths a factor of 2 and 4 larger, respectively. The horizontal line shows the typical Automatic Plate Measuring (APM) machine isophotal threshold (Loveday et al. 1992). Galaxy C would be missed entirely by the survey, while galaxy B would have a measured "total" magnitude too faint by 0.7 mag, if the constant isophotal-to-total magnitude correction were based on galaxy A. The horizontal dashed line shows the isophotal threshold for the Fornax cluster survey (Ferguson 1989). The right panel shows a simulated field galaxy sample made up of exponential-profile galaxies with a steep intrinsic surface-brightness–luminosity relation and a luminosity function with $M^*_{B_J} = -21$, $\alpha = -1.5$ in the $B_J$ band (see text). If galaxies could be detected independent of surface brightness and their true total magnitudes measured, the intrinsic luminosity function would be recovered (open triangles). However, a survey that selects galaxies only above an isophotal threshold of $\mu = 24.5$, and adopts magnitudes based on a constant isophotal-to-total magnitude correction ($\mathrm{m_{tot}} = \mathrm{m_{iso}} - 0.27$ assumed here), would recover the luminosity function shown by the solid circles, similar to that observed by Loveday *et al.* (1992). Further details of such simulations can be found in Ferguson & McGaugh (1994).

Figure 8:
The projected density of galaxies as a function of radius in the Virgo cluster. The solid line shows the best-fit exponential to the bright non-nucleated dE distribution; the dotted line shows the best fit to the faint non-nucleated dE distribution. Note how the *bright* non-nucleated dE's follow the distribution of late-type cluster members. The effect is also seen in the Fornax cluster. From Ferguson & Sandage 1989.